\documentclass[10pt,aps,prx,twocolumn,notitlepage,showpacs,superscriptaddress]{revtex4-1}
\usepackage{amsthm,amsmath,amsfonts,amssymb,verbatim,color}
\usepackage{graphicx}
\usepackage{bm}
\usepackage{epsfig,slashed}
\usepackage{amssymb}
\usepackage{amsfonts}
\usepackage{tikz}
\usepackage{xparse}
\usepackage{ifthen}
\usepackage{lipsum}
\usepackage[colorlinks=true,citecolor=blue,
linkcolor=blue,urlcolor=blue]{hyperref}
\usepackage[caption=false]{subfig}
\begin{document}

\newcommand{\tr}{\mathop{\mathrm{tr}}}
\newcommand{\bsigma}{\boldsymbol{\sigma}}
\newcommand{\re}{\mathop{\mathrm{Re}}}
\newcommand{\im}{\mathop{\mathrm{Im}}}
\renewcommand{\b}[1]{{\boldsymbol{#1}}}
\newcommand{\diag}{\mathrm{diag}}
\newcommand{\sign}{\mathrm{sign}}
\newcommand{\sgn}{\mathop{\mathrm{sgn}}}

\newcommand{\cl}{\mathrm{cl}}
\newcommand{\mb}{\bm}
\newcommand{\ua}{\uparrow}
\newcommand{\da}{\downarrow}
\newcommand{\ra}{\rightarrow}
\newcommand{\la}{\leftarrow}
\newcommand{\mc}{\mathcal}
\newcommand{\bs}{\boldsymbol}
\newcommand{\lra}{\leftrightarrow}
\newcommand{\nn}{\nonumber}
\newcommand{\half}{{\textstyle{\frac{1}{2}}}}
\newcommand{\mf}{\mathfrak}
\newcommand{\MF}{\text{MF}}
\newcommand{\IR}{\text{IR}}
\newcommand{\UV}{\text{UV}}
\newcommand{\sech}{\mathrm{sech}}

\title{Disordered fermionic quantum critical points}

\author{Hennadii Yerzhakov}
\affiliation{Department of Physics, University of Alberta, Edmonton, Alberta T6G 2E1, Canada}

\author{Joseph Maciejko}
\affiliation{Department of Physics, University of Alberta, Edmonton, Alberta T6G 2E1, Canada}
\affiliation{Theoretical Physics Institute, University of Alberta, Edmonton, Alberta T6G 2E1, Canada}
\affiliation{Canadian Institute for Advanced Research, Toronto, Ontario M5G 1Z8, Canada}

\date\today

\begin{abstract}
We study the effect of quenched disorder on the semimetal-superconductor quantum phase transition in a model of two-dimensional Dirac semimetal with $N$ flavors of two-component Dirac fermions, using perturbative renormalization group methods at one-loop order in a double epsilon expansion. For $N\geq 2$ we find that the Harris-stable clean critical behavior gives way, past a certain critical disorder strength, to a finite-disorder critical point characterized by non-Gaussian critical exponents, a noninteger dynamic critical exponent $z>1$, and a finite Yukawa coupling between Dirac fermions and bosonic order parameter fluctuations. For $N\geq 7$ the disordered quantum critical point is described by a renormalization group fixed point of stable-focus type and exhibits oscillatory corrections to scaling.
\end{abstract}

\maketitle

\section{Introduction}

The study of Dirac fermions in the presence of quenched disorder is a problem of enduring interest due to its relevance for a remarkable breadth of phenomena in condensed matter physics, with early applications including disordered zero-gap semiconductors~\cite{fradkin1986,fradkin1986b}, the random-bond Ising model~\cite{shankar1987}, and the integer quantum Hall plateau transition~\cite{ludwig1994}. The discovery of three-dimensional (3D) topological semimetals~\cite{TopoSM} has led to renewed interest in this problem, as evidenced by the large number of theoretical studies of disordered Weyl~\cite{huang2013,sbierski2014,syzranov2015,ominato2015,zhao2015,altland2015,sbierski2015,chen2015,bera2016,roy2016,louvet2016,pixley2017,holder2017,wilson2018,roy2016b,syzranov2018} and Dirac~\cite{kobayashi2014,nandkishore2014,skinner2014,roy2014,pixley2015,pixley2016} semimetals having appeared in recent years. While this body of work has largely focused on the noninteracting limit, relatively fewer studies have addressed the combined effect of disorder and electron-electron interactions in Dirac fermion systems. Limiting ourselves to 2D Dirac fermions, our prime concern, such studies have addressed the interplay of interactions and disorder on the integer quantum Hall plateau transition~\cite{ye1999}, the physics of graphene~\cite{stauber2005,herbut2008,vafek2008,foster2008,wang2014,potirniche2014}, and the surfaces of 3D topological insulators~\cite{nandkishore2013,konig2013,ozfidan2016} and superconductors~\cite{foster2012,foster2014,ghorashi2017}. Recent work has also demonstrated the possibility of novel critical phases in massless (2+1)D relativistic quantum electrodynamics in the presence of quenched disorder~\cite{goswami2017,thomson2017,zhao2017}, with possible applications to disordered spin liquids.

In this work we study the effect of quenched disorder on the semimetal-superconductor quantum phase transition of 2D Dirac fermions at charge neutrality. While previous work involving one of us has already partially addressed this problem using mean-field~\cite{nandkishore2013,potirniche2014} and standard epsilon expansion~\cite{nandkishore2013} methods, here we revisit this problem using the double epsilon expansion~\cite{dorogovtsev1980,boyanovsky1982,lawrie1984} which is better suited to the study of quantum critical phenomena in disordered systems. While the double epsilon expansion has traditionally been applied to purely bosonic systems, e.g., the $O(n)$ vector model with random-$T_c$ disorder~\cite{dorogovtsev1980,boyanovsky1982,lawrie1984}, here we show that it can be applied to fermionic quantum critical points (QCPs) described by quantum field theories of the Gross-Neveu-Yukawa (GNY) type~\cite{zinn-justin1991,rosenstein1993}, exploiting the fact that, like the $O(n)$ vector model, such theories have an upper critical dimension of four absent quenched disorder. We consider a model of 2D Dirac semimetal with $N$ flavors of two-component Dirac fermions, and show that at leading (one-loop) order in the double epsilon expansion, a Harris-stable clean QCP gives way beyond a certain critical disorder strength to a finite-disorder QCP~\cite{motrunich2000} with non-Gaussian critical exponents and noninteger dynamic critical exponent $z>1$. Furthermore, Dirac fermions and bosonic order parameter fluctuations are strongly coupled at this QCP. The latter is therefore a first example of \emph{disordered fermionic QCP}, which combines the phenomenology of finite-disorder bosonic QCPs~\cite{vojta2013} with that of (clean) fermionic QCPs, where coupling between bosonic order parameter fluctuations and gapless fermionic modes leads to new universality classes beyond those of the purely bosonic Landau-Ginzburg-Wilson paradigm.

The paper is structured as follows. In Sec.~\ref{sec:model} we present our model for the semimetal-superconductor transition in the presence of quenched disorder. In Sec.~\ref{sec:RG} we outline the basic steps of the renormalization group (RG) approach in the double epsilon expansion and present the beta functions describing the flow under renormalization of various coupling constants in the theory. In Sec.~\ref{sec:RGflowanalysis} we find RG fixed points, analyze their stability, and determine how they are connected under the RG flow. In Sec.~\ref{sec:critical} we determine the critical exponents at the various fixed points and derive implications of the RG flow analysis for the phase diagram of the system. A brief conclusion follows in Sec.~\ref{sec:conclusion}, and the details of some derivations are contained in two appendices to the paper.

\section{Model}
\label{sec:model}

We consider a model of $N$ flavors of two-component Dirac fermions $\psi^1,\psi^2,\ldots,\psi^N$ in 2+1 dimensions, which in the absence of interactions are described by the low-energy imaginary-time Lagrangian
\begin{align}\label{LpsiClean}
\mathcal{L}_\psi=\sum_{i=1}^Ni\bar{\psi}^i(\gamma_0\partial_\tau+c_f\boldsymbol{\gamma}\cdot\nabla)\psi^i,
\end{align}
where $\gamma_0$ and $\boldsymbol{\gamma}=(\gamma_1,\gamma_2)$ denote Euclidean $2\times 2$ Dirac matrices in 2+1 dimensions, obeying the $SO(3)$ Clifford algebra $\{\gamma_\mu,\gamma_\nu\}=2\delta_{\mu\nu}\mathbb{I}_{2\times 2}$, $\mu,\nu=0,1,2$, with $\mathbb{I}_{2\times 2}$ the $2\times 2$ identity matrix, and $\bar{\psi}^i=-i\psi^{i\dag}\gamma_0$ is the Dirac conjugate. In a condensed matter system on a lattice the $N$ flavors would correspond to $N$ symmetry-related linear band crossings in the Brillouin zone, with a common Dirac velocity $c_f$. We also assume the underlying microscopic model is particle-hole symmetric, which excludes any possible tilt of the Dirac cones. For a 3D topological insulator the two components of the spinor $\psi^i$ correspond to physical spin; for a 2D Dirac semimetal like graphene an equivalent four-component formulation is more natural (see Appendix~\ref{app:spinors}).

We will be interested in superconducting instabilities, and consider subjecting the Dirac fermions to sufficiently short-range attractive interactions. At low energies, the various possible superconducting order parameters will transform according to irreducible representations of the symmetry group of (\ref{LpsiClean}). We will assume the microscopic interactions are such that in a certain range of couplings they favor pairing in the flavor-symmetric, $s$-wave, spin-singlet channel, with an order parameter
\begin{align}
\sum_{i=1}^N\langle\psi^{iT}i\sigma_2\psi^i\rangle,
\end{align}
where $T$ denotes the transpose and $\sigma_1,\sigma_2,\sigma_3$ are the Pauli spin matrices, which act on the physical spin degrees of freedom. We consider first the clean limit, and assume that the chemical potential is exactly at the Dirac point. The transition from Dirac semimetal to superconductor at zero temperature proceeds via a QCP at finite attraction strength, since the density of states of the Dirac semimetal vanishes at the Fermi energy~\cite{wilson1973,gross1974,zhao2006,kopnin2008,roy2013,nandkishore2013}. The critical behavior at the QCP is governed by the so-called chiral XY GNY model~\cite{rosenstein1993},
\begin{align}\label{Lclean}
\mathcal{L}_\text{clean}=\mathcal{L}_\psi+\mathcal{L}_\phi+\mathcal{L}_{\phi\psi\psi},
\end{align}
where
\begin{align}
\mathcal{L}_\phi&=|\partial_\tau\phi|^2+c_b^2|\nabla\phi|^2+r|\phi|^2+\lambda^2|\phi|^4,\label{Lphi}\\
\mathcal{L}_{\phi\psi\psi}&=h\phi^*\sum_{i=1}^N\psi^{iT}i\sigma_2\psi^i+\text{H.c.}
\end{align}
The Lagrangian (\ref{Lclean}) describes gapless Dirac fermions interacting with bosonic order parameter fluctuations $\phi$ with velocity $c_b$; $r$ is a tuning parameter for the transition ($r>0$ in the semimetal phase, $r<0$ in the superconducting phase, and $r=0$ at criticality), and the coupling constants $\lambda^2$ and $h$ obey $\lambda^2>0$ and $h^2>0$. The absence of a term $\phi^*\partial_\tau\phi$ linear in time derivatives is a consequence of the assumed particle-hole symmetry of the underlying microscopic model. The effective low-energy Lagrangian (\ref{Lclean}) exhibits an emergent $O(N)$ flavor symmetry under $\psi^i\rightarrow W_{ij}\psi^j$, with $W$ an arbitrary orthogonal $N\times N$ matrix, and its critical properties for any $N$ can be accessed via an RG analysis in $D=4-\epsilon$ spacetime dimensions~\cite{rosenstein1993,roy2013,zerf2017,boyack2018}. For $N=1$, the model is applicable to the superconducting transition on the surface of a 3D topological insulator with a single Dirac cone, and features a QCP with emergent $\mathcal{N}=2$ supersymmetry~\cite{balents1998,ScottThomas,lee2007,grover2014,ponte2014,zerf2016,fei2016,zerf2017,li2017b}. For $N=4$ the model describes the superconducting transition in graphene~\cite{roy2013}. In the infrared limit, in which a $\mathbb{Z}_3$ anisotropy $\sim(\phi^3+\phi^{*3})$ becomes irrelevant, the $N=4$ case is argued to also belong to the same universality class as that of the Kekul\'e valence-bond-solid transition in monolayer graphene~\cite{hou2007,ryu2009,roy2010,zhou2016,li2017,scherer2016,classen2017}, and possibly also twisted bilayer graphene~\cite{xu2018}. In Appendix~\ref{app:spinors} we establish an equivalence between the two-component formulation with Yukawa coupling to the Majorana mass used here and in Ref.~\cite{zerf2016}, and a four-component formulation with normal and axial Dirac masses typically used in discussions of graphene~\cite{roy2013,zerf2017}, where the $U(1)$ symmetry is realized as an axial symmetry.

\begin{figure}[t]
\includegraphics[width=0.5\columnwidth]{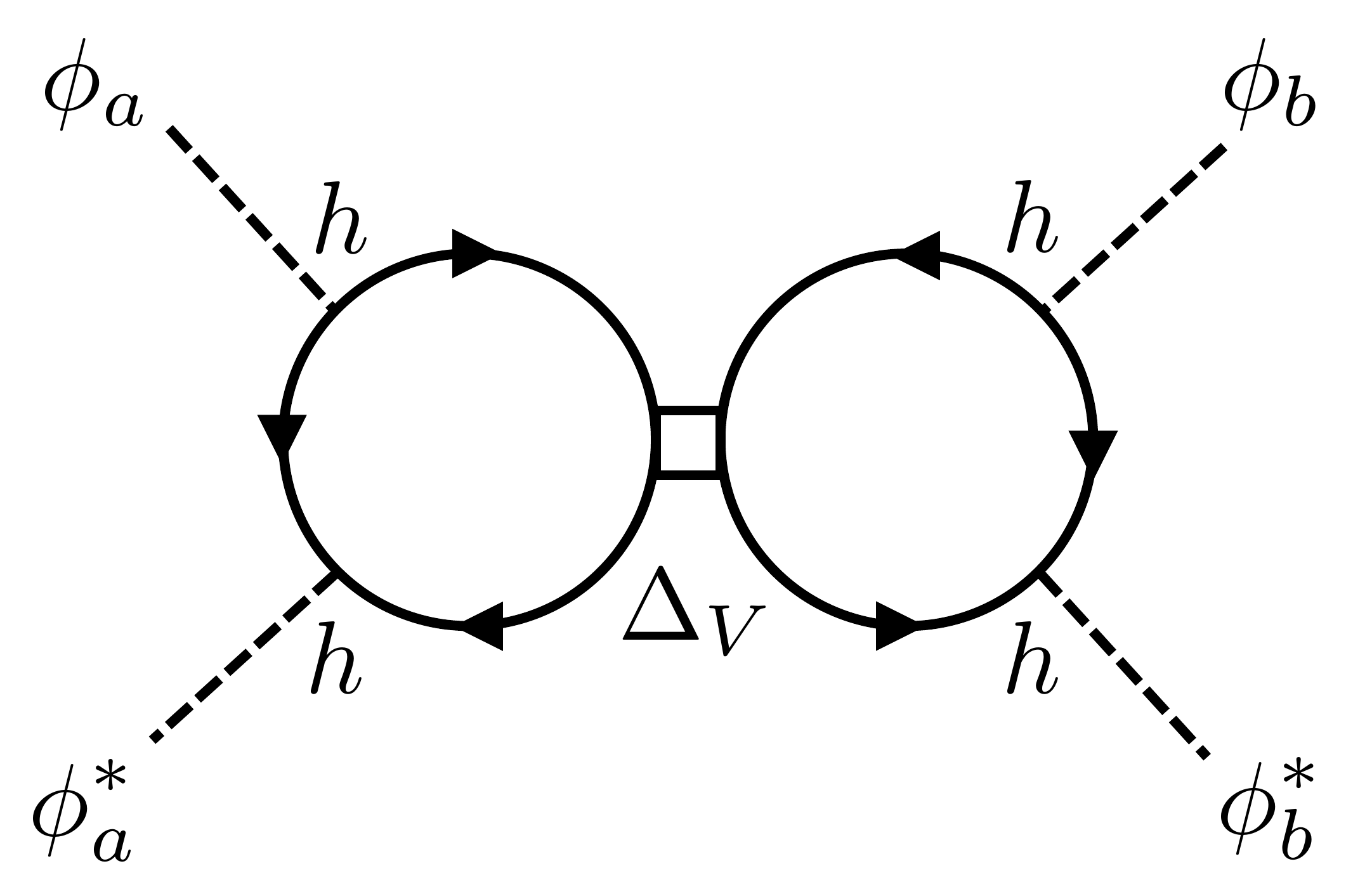}
\caption{Random-$T_c$ disorder is generated from random chemical potential disorder at two-loop order (dotted lines: order parameter fluctuations, solid lines: fermions, box: disorder-induced four-fermion coupling).}
\label{fig:2loop}
\end{figure}

Focusing on the superconducting transition, we now consider the effect of quenched disorder on this transition. We assume a random potential $V(\b{x})$ that is smooth on the scale of the microscopic lattice constant, i.e., that is sufficiently long-range so as to not scatter Dirac fermions between different valleys (see, e.g., Ref.~\cite{ando1998}). The potential then couples identically to all fermion flavors,
\begin{align}
\mathcal{L}_\text{dis}=V(\b{x})\sum_{i=1}^N\psi^{i\dag}\psi^i.
\end{align}
Proceeding as in Ref.~\cite{nandkishore2013}, we assume a Gaussian disorder distribution with zero mean and variance $\Delta_V$,
\begin{align}
P[V(\b{x})]\propto e^{-\int d^2\b{x}\,V(\b{x})^2/2\Delta_V},
\end{align}
and perform the quenched disorder average using the replica trick~\cite{QPT}. This generates a four-fermion interaction nonlocal in time,
\begin{align}\label{SdisMu}
S_\text{dis,f}=-\frac{\Delta_V}{2}\sum_{a,b=1}^n\sum_{i,j=1}^N\int d^2\b{x}\,&d\tau\,d\tau'\,(\psi^{i\dag}_a\psi^i_a)(\b{x},\tau)\nn\\
&\times(\psi^{j\dag}_b\psi^j_b)(\b{x},\tau'),
\end{align}
where the replica limit $n\rightarrow 0$ is to be taken at the end of the calculation. This effective interaction preserves all the symmetries of the clean limit, including translation symmetry and $O(N)$ flavor symmetry. As will be explained in greater detail in Sec.~\ref{sec:RG}, in the context of an RG analysis near four dimensions the four-fermion interaction term (\ref{SdisMu}) is strongly irrelevant in perturbation theory, and thus would not appear to affect critical behavior in the scaling limit. However, at two-loop order this interaction generates an effective four-boson interaction,
\begin{align}\label{SdisBos}
S_\text{dis,b}=-\frac{\Delta}{2}\sum_{a,b=1}^n\int d^2\b{x}\,&d\tau\,d\tau'\,|\phi_a|^2(\b{x},\tau)|\phi_b|^2(\b{x},\tau'),
\end{align}
where $\Delta\propto h^4\Delta_V$ at leading order in perturbation theory (Fig.~\ref{fig:2loop}). The four-boson interaction (\ref{SdisBos}) is identical to one generated by Gaussian disorder in the coefficient of the $|\phi|^2$ term in Eq.~(\ref{Lphi}), i.e., random-$T_c$ disorder. By contrast with Eq.~(\ref{SdisMu}), this interaction is relevant below four dimensions~\cite{lubensky1975} and must be included in an RG analysis of the critical behavior, to which we now turn.

\section{RG in the double epsilon expansion}
\label{sec:RG}

In the limit of a unique fermion flavor $N=1$, the problem so far described has been studied in Ref.~\cite{nandkishore2013} using the $\epsilon$ expansion in $D=4-\epsilon$ spacetime dimensions. In this expansion the four-fermion coupling $\Delta_V$ in Eq.~(\ref{SdisMu}) has an engineering dimension $-1+\epsilon$, and is thus strongly irrelevant at the Gaussian fixed point for small $\epsilon$, while the induced four-boson coupling $\Delta$ in Eq.~(\ref{SdisBos}) has an engineering dimension $1+\epsilon$, which is strongly relevant at the Gaussian fixed point. In the $\epsilon$ expansion one thus finds that disorder is relevant at the clean QCP also~\cite{nandkishore2013}, since dimensions of operators at this QCP only receive $\mathcal{O}(\epsilon)$ corrections relative to their engineering dimensions. In fact, the conventional $\epsilon$ expansion below four dimensions generally predicts runaway flows near QCPs with random-$T_c$ disorder~\cite{QPT}. While such runaway flows are often interpreted as an indication that critical behavior is destroyed, they really only signal the breakdown of the conventional $\epsilon$ expansion as well as the need for another small parameter with which to tame RG flows generated by disorder. Here we will follow one particular approach to fulfill this need, which consists in working in $d=4-\epsilon$ spatial and $\epsilon_\tau$ time dimensions, with both $\epsilon$ and $\epsilon_\tau$ treated as small parameters~\cite{dorogovtsev1980,boyanovsky1982,lawrie1984}. In the present case, to access the physical problem in 2+1 dimensions one extrapolates $\epsilon\rightarrow 2$ and $\epsilon_\tau\rightarrow 1$. (For a study of quantum critical phenomena in disordered 3D Dirac semimetals using a different type of double epsilon expansion, see Ref.~\cite{roy2016c}.)

\subsection{Bare vs renormalized actions}

Focusing first on the critical theory $r=0$, we thus study the replicated action
\begin{widetext}
\begin{align}\label{Sreplica}
S=&\sum_a\int d^d\b{x}\,d^{\epsilon_\tau}\tau\biggl(i\bar{\psi}_a(\slashed{\partial}_\tau+c_f\slashed{\nabla})\psi_a+|\partial_\tau\phi_a|^2+c_b^2|\nabla\phi_a|^2+\lambda^2|\phi_a|^4
+h(\phi^*_a\psi_a^Ti\sigma_2\psi_a+\text{H.c.})\biggr)\nn\\
&-\frac{\Delta}{2}\sum_{ab}\int d^d\b{x}\,d^{\epsilon_\tau}\tau\,d^{\epsilon_\tau}\tau'|\phi_a|^2(\b{x},\tau)|\phi_b|^2(\b{x},\tau'),
\end{align}
where $a,b=1,\ldots,n$ are replica indices, we denote $\slashed{\partial}_\tau\equiv\gamma_0\partial_\tau$ and $\slashed{\nabla}\equiv\boldsymbol{\gamma}\cdot\nabla$ for simplicity, and we group the $N$ fermion flavors for each replica $a$ into an $O(N)$ vector, $\psi_a\equiv(\psi_a^1,\psi_a^2,\ldots,\psi_a^N)$. By rescaling the fermion and boson fields as well as the time coordinate, and redefining the couplings in the Lagrangian, one can eliminate the velocities $c_f$ and $c_b$ from the Lagrangian at the expense of multiplying $|\partial_\tau\phi_a|^2$ by the ratio $(c_f/c_b)^2$, which we will denote $c^2$.

To carry out an RG analysis of the above theory, we compare the bare action
\begin{align}
S_B=&\sum_a\int d^d\b{x}_B\,d^{\epsilon_\tau}\tau_B\biggl(i\bar{\psi}_{a,B}(\slashed{\partial}_{\tau_B}+\slashed{\nabla}_B)\psi_{a,B}
+c^2_B|\partial_{\tau_B}\phi_{a,B}|^2+|\nabla_B\phi_{a,B}|^2+\lambda_B^2|\phi_{a,B}|^4\nn\\
&+h_B(\phi_{a,B}^*\psi_{a,B}^Ti\sigma_2\psi_{a,B}+\text{H.c.})\biggr)
-\frac{\Delta_B}{2}\sum_{ab}\int d^d\b{x}_B\,d^{\epsilon_\tau}\tau_B\,d^{\epsilon_\tau}\tau_B'
|\phi_{a,B}|^2(\b{x}_B,\tau_B)|\phi_{b,B}|^2(\b{x}_B,\tau_B'),
\end{align}
to the renormalized action
\begin{align}\label{Sren}
S=&\sum_a\int d^d\b{x}\,d^{\epsilon_\tau}\tau\biggl(Z_1i\bar{\psi}_a\slashed{\partial}_\tau\psi_a
+Z_2i\bar{\psi}_a\slashed{\nabla}\psi_a+Z_3c^2|\partial_\tau\phi_a|^2+Z_4|\nabla\phi_a|^2+Z_5\lambda^2\mu^{\epsilon-\epsilon_\tau}|\phi_a|^4\nn\\
&+Z_6h\mu^{(\epsilon-\epsilon_\tau)/2}(\phi_a^*\psi_a^Ti\sigma_2\psi_a+\text{H.c.})\biggr)
-Z_7\frac{\Delta}{2}\mu^\epsilon\sum_{ab}\int d^d\b{x}\,d^{\epsilon_\tau}\tau\,d^{\epsilon_\tau}\tau'\,
|\phi_a|^2(\b{x},\tau)|\phi_b|^2(\b{x},\tau'),
\end{align}
\end{widetext}
where the renormalized couplings $c^2$, $\lambda^2$, $h$, $\Delta$ are dimensionless, and we have introduced a renormalization scale $\mu$. The renormalization constants $Z_1,\ldots,Z_7$ are to be calculated in perturbation theory. The bare and renormalized kinetic terms for the fermion match if one takes $\b{x}_B=\b{x}$, $\tau_B=\eta\tau$, and
\begin{align}
\sqrt{Z_1}\psi_a(\b{x},\tau)&=\eta^{(\epsilon_\tau-1)/2}\psi_{a,B}(\b{x}_B,\tau_B),\label{FermionZ1}\\
\sqrt{Z_2}\psi_a(\b{x},\tau)&=\eta^{\epsilon_\tau/2}\psi_{a,B}(\b{x}_B,\tau_B),\label{FermionZ2}
\end{align}
which implies $\eta=Z_2/Z_1$. The dynamic critical exponent $z$ describes the relative scaling of space and time, which in dimensionless units reads $\mu\tau\sim(\mu|\b{x}|)^z$. Defining the anomalous dimensions
\begin{align}
\gamma_i=\frac{d\ln Z_i}{d\ln\mu},\,i=1,\ldots,7,
\end{align}
this implies~\cite{thomson2017}
\begin{align}
z=1+\gamma_1-\gamma_2,
\end{align}
since the bare coordinate $\b{x}_B$ and time $\tau_B$ do not depend on $\mu$. Likewise, the $|\nabla\phi|^2$ terms match if one requires
\begin{align}
\sqrt{Z_4}\phi_a(\b{x},\tau)=\eta^{\epsilon_\tau/2}\phi_{a,B}(\b{x}_B,\tau_B).\label{BosonZ4}
\end{align}

From Eq.~(\ref{FermionZ1})-(\ref{FermionZ2}) and (\ref{BosonZ4}) we find that the bare and renormalized coupling constants are related by
\begin{align}
c^2&=Z_3^{-1}Z_4\left(\frac{Z_1}{Z_2}\right)^2 c_B^2,\\
\lambda^2&=\mu^{-(\epsilon-\epsilon_\tau)}\left(\frac{Z_1}{Z_2}\right)^{\epsilon_\tau}Z_4^2Z_5^{-1}\lambda_B^2,\\
h^2&=\mu^{-(\epsilon-\epsilon_\tau)}\left(\frac{Z_1}{Z_2}\right)^{\epsilon_\tau}Z_2^2Z_4Z_6^{-2}h_B^2,\\
\Delta&=\mu^{-\epsilon}Z_4^2Z_7^{-1}\Delta_B,
\end{align}
from which we obtain the RG beta functions $\beta_g\equiv dg/d\ln \mu$, $g\in\{c^2,\lambda^2,h^2,\Delta\}$,
\begin{align}
\beta_{c^2}&=(2\gamma_1-2\gamma_2-\gamma_3+\gamma_4)c^2,\label{betac2}\\
\beta_{\lambda^2}&=\bigl(-(\epsilon-\epsilon_\tau)+\epsilon_\tau(\gamma_1-\gamma_2)+2\gamma_4-\gamma_5\bigr)
\lambda^2,\\
\beta_{h^2}&=\bigl(-(\epsilon-\epsilon_\tau)+\epsilon_\tau(\gamma_1-\gamma_2)+2\gamma_2+\gamma_4-2\gamma_6\bigr)h^2,\\
\beta_\Delta&=(-\epsilon+2\gamma_4-\gamma_7)\Delta,\label{betaDelta}
\end{align}
using the fact that the bare couplings $c_B^2$, $\lambda_B^2$, $h_B^2$, and $\Delta_B$ are independent of $\mu$. For $\epsilon>\epsilon_\tau>0$, the couplings $\lambda^2$, $h^2$, and $\Delta$ are relevant at the Gaussian fixed point, and one may hope to find a controlled fixed point in perturbation theory for small $\epsilon,\epsilon_\tau$. Note that at tree level, the fermion field has scaling dimension $[\psi]=(3-\epsilon+\epsilon_\tau)/2$ and the boson field, $[\phi]=(2-\epsilon+\epsilon_\tau)/2$. Therefore the four-fermion disorder-induced coupling $\Delta_V$ in Eq.~(\ref{SdisMu}) has dimension
\begin{align}
d+2\epsilon_\tau-[\psi^\dag\psi\psi^\dag\psi]=-2+\epsilon,
\end{align}
which is strongly irrelevant for small $\epsilon,\epsilon_\tau$, justifying our excluding it from the action (\ref{Sreplica}).

To determine the correlation length exponent $\nu$ one needs to compute the RG eigenvalue of the scalar field mass term $|\phi|^2$ at the QCP, which is done by adding the term $\sum_a r_B|\phi_{a,B}|^2$ to the bare Lagrangian and $\sum_a Z_rr\mu^2|\phi_a|^2$ to its renormalized counterpart. Equating the two gives the relation
\begin{align}
r=\mu^{-2}Z_4Z_r^{-1}r_B,
\end{align}
which yields the usual expression for the inverse correlation length exponent~\cite{ZJ},
\begin{align}
\nu^{-1}=2-\gamma_4+\gamma_r,
\end{align}
defining $\gamma_r=d\ln Z_r/d\ln\mu$ as for the other renormalization constants. Finally, the fermion $\gamma_\psi$ and boson $\gamma_\phi$ anomalous dimensions are obtained from $\gamma_{\psi,\phi}=d\ln Z_{\psi,\phi}/d\ln\mu$ where we define $Z_\psi$ and $Z_\phi$ via
\begin{align}
\psi_{a,B}(\b{x}_B,\tau_B)&=\sqrt{Z_\psi}\psi_a(\b{x},\tau),\\
\phi_{a,B}(\b{x}_B,\tau_B)&=\sqrt{Z_\phi}\phi_a(\b{x},\tau).
\end{align}
Using Eq.~(\ref{FermionZ1})-(\ref{FermionZ2}) and (\ref{BosonZ4}) we find
\begin{align}
\gamma_\psi&=\gamma_2+\epsilon_\tau(z-1),\label{gammapsi1}\\
\gamma_\phi&=\gamma_4+\epsilon_\tau(z-1).\label{gammaphi1}
\end{align}

\subsection{Renormalization constants}

\begin{figure}[t]
\includegraphics[width=\columnwidth]{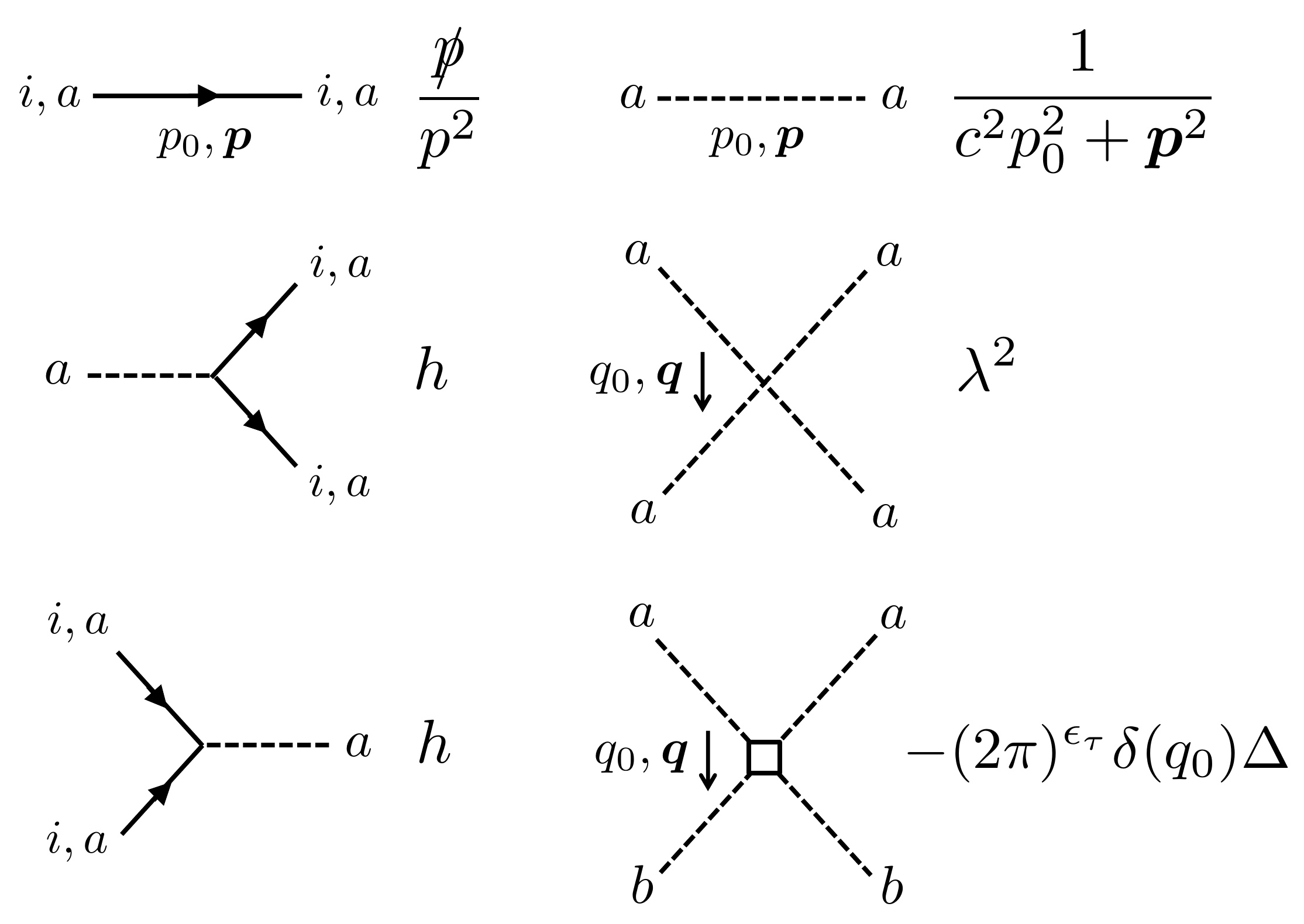}
\caption{Feynman rules associated with the replicated action; $a,b$ are replica indices, $i$ is a fermion flavor index, and $q_0,\b{q}$ denotes the frequency-momentum transfer from top to bottom.}
\label{fig:feynrules}
\end{figure}

To derive the beta functions (\ref{betac2})-(\ref{betaDelta}) one must first compute the renormalization constants $Z_1,\ldots,Z_7$, and to determine the correlation length exponent one must calculate $Z_r$. Here we adopt the standard field-theoretic approach, with renormalization constants calculated at one-loop order in the modified minimal subtraction ($\overline{\text{MS}}$) scheme with dimensional regularization. The Feynman rules associated with the replicated action are illustrated schematically in Fig.~\ref{fig:feynrules}; the fermion and boson propagators are given by
\begin{align}
G_{ab}^{ij}(p)&=\langle\psi_a^i(p)\bar{\psi}_b^j(p)\rangle=\delta_{ab}\delta^{ij}\frac{\slashed{p}}{p^2},\\
D_{ab}(p)&=\langle\phi_a(p)\phi^*_b(p)\rangle=\frac{\delta_{ab}}{c^2p_0^2+\b{p}^2},
\end{align}
denoting the spacetime momentum by $p=(p_0,\b{p})$ and $\slashed{p}=\gamma_\mu p_\mu$.

\begin{figure}[t]
\includegraphics[width=\columnwidth]{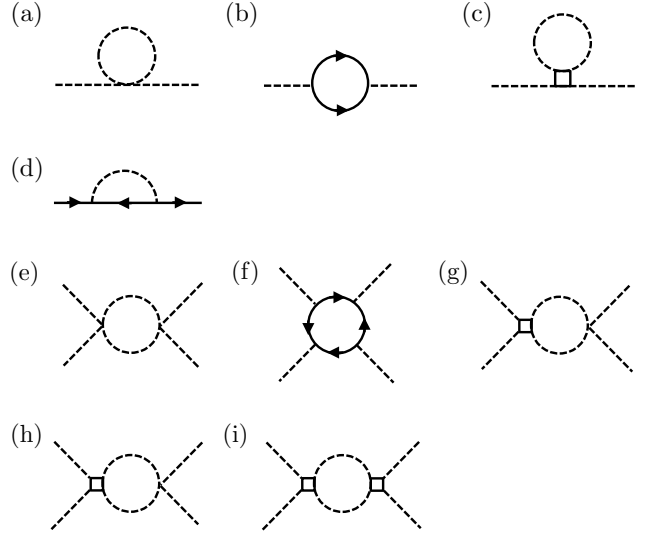}
\caption{One-loop diagrams for the renormalization of (a,b,c) the boson two-point function; (d) the fermion two-point function; (e,f,g) the boson self-interaction $\lambda^2$; (h,i) the disorder strength $\Delta$. At this order there is no renormalization of the Yukawa coupling $h$.}
\label{fig:diagrams}
\end{figure}

In the $\overline{\text{MS}}$ scheme, the renormalization constants are computed order by order in the loop expansion by writing $Z_i=1+\delta Z_i$, $i=1,\ldots,7,r$ and demanding that the $\delta Z_i$ cancel the ultraviolet divergences of the one-particle irreducible (1PI) effective action. In dimensional regularization, this means that at one-loop order the $\delta Z_i$, which are computed from the Feynman diagrams in Fig.~\ref{fig:diagrams}, contain simple poles in $\epsilon$ and $\epsilon-\epsilon_\tau$. We present the details of the calculation in Appendix~\ref{app:Z}; here we simply quote the results (after taking the replica limit $n\rightarrow 0$):
\begin{align}
Z_1&=1-\frac{8h^2}{\epsilon-\epsilon_\tau}f(c^2),\label{Z1}\\
Z_2&=1-\frac{4h^2}{\epsilon-\epsilon_\tau},\\
Z_3&=1-\frac{2\Delta}{\epsilon}-\frac{4Nh^2c^{-2}}{\epsilon-\epsilon_\tau},\\
Z_4&=1-\frac{4Nh^2}{\epsilon-\epsilon_\tau},\\
Z_5&=1+\frac{20\lambda^2}{\epsilon-\epsilon_\tau}-\frac{16Nh^4\lambda^{-2}}{\epsilon-\epsilon_\tau}-\frac{12\Delta}{\epsilon},\\
Z_6&=1,\\
Z_7&=1+\frac{16\lambda^2}{\epsilon-\epsilon_\tau}-\frac{8\Delta}{\epsilon},\label{Z7}\\
Z_r&=1+\frac{8\lambda^2}{\epsilon-\epsilon_\tau}-\frac{2\Delta}{\epsilon},\label{Zr}
\end{align}
where we have rescaled the coupling constants according to $g/(4\pi)^2\rightarrow g$, $g\in\{\lambda^2,h^2,\Delta\}$, and we define the dimensionless function (see Fig.~\ref{fig:f}),
\begin{align}\label{f}
f(c^2)=\frac{c^2(c^2-1-\ln c^2)}{(c^2-1)^2}.
\end{align}

\subsection{Beta functions and anomalous dimensions}

To calculate the beta functions, we first use the chain rule to write
\begin{align}\label{gammai}
\gamma_i=\frac{1}{Z_i}\frac{dZ_i}{d\ln\mu}=\frac{1}{Z_i}\sum_g\frac{\partial Z_i}{\partial g}\beta_g,
\end{align}
for $i=1,\ldots,7$ and $g\in\{c^2,\lambda^2,h^2,\Delta\}$, which when substituted into the expressions (\ref{betac2})-(\ref{betaDelta}) gives a linear system of equations for the beta functions. Expanding the beta functions to quadratic order in the couplings, we find that all poles in $\epsilon$ and $\epsilon-\epsilon_\tau$ cancel, and obtain
\begin{align}
\beta_{c^2}&=-2c^2\Delta+4h^2\left[c^2\left(4f(c^2)+N-2\right)-N\right],\label{bc2}\\
\beta_{\lambda^2}&=-(\epsilon-\epsilon_\tau)\lambda^2-12\Delta\lambda^2+20\lambda^4+8Nh^2\lambda^2-16Nh^4,\label{bl2}\\
\beta_{h^2}&=-(\epsilon-\epsilon_\tau)h^2+4(N+2)h^4,\label{bh2}\\
\beta_\Delta&=-\epsilon\Delta-8\Delta^2+16\Delta\lambda^2+8N\Delta h^2.\label{bD}
\end{align}
Setting $\epsilon_\tau=0$ and $\Delta=0$, Eq.~(\ref{bh2}) and (\ref{bl2}) reduce to the one-loop beta functions of the chiral XY GNY model in the ordinary $4-\epsilon$ expansion (e.g., Eq.~(19)-(20) in Ref.~\cite{boyack2018} in the $e^2=0$ limit). Note that the above beta functions are perturbative in $\lambda^2$, $h^2$, and $\Delta$, but exact in the relative velocity parameter $c^2$.

Using Eq.~(\ref{gammai}), from the renormalization constants (\ref{Z1})-(\ref{Z7}) and the beta functions (\ref{bc2})-(\ref{bD}) we can calculate the anomalous dimensions $\gamma_i$, and from those the critical exponents $\nu^{-1}$, $z$, $\gamma_\psi$, and $\gamma_\phi$. We obtain
\begin{align}
\nu^{-1}&=2-4Nh^2-8\lambda^2+2\Delta,\label{nuinv}\\
z&=1+4h^2\bigl(2f(c^2)-1\bigr),\label{z}\\
\gamma_\psi&=4h^2\left[1+\bigl(2f(c^2)-1\bigr)\epsilon_\tau\right],\label{gammapsi}\\
\gamma_\phi&=4Nh^2\left[1+\bigl(2f(c^2)-1\bigr)\frac{\epsilon_\tau}{N}\right],\label{gammaphi}
\end{align}
which are meant to be evaluated at the RG fixed points $(c^2_*,\lambda^2_*,h^2_*,\Delta_*)$ discussed in the following section. At one-loop order $h_*^2\sim\mathcal{O}(\epsilon,\epsilon_\tau)$, thus the subleading correction proportional to $\epsilon_\tau$ in the fermion (\ref{gammapsi}) and boson (\ref{gammaphi}) anomalous dimensions should be discarded. In other words, at one-loop order the correction $z-1$ to the dynamic critical exponent is $\mathcal{O}(\epsilon,\epsilon_\tau)$, which gives a term quadratic in $\epsilon,\epsilon_\tau$ in Eq.~(\ref{gammapsi1})-(\ref{gammaphi1}) that should be treated on par with two-loop corrections to $\gamma_2$, $\gamma_4$, and thus eliminated when working at one-loop order.

\section{RG flow analysis}
\label{sec:RGflowanalysis}

We now search for fixed points of the flow equations (\ref{bc2})-(\ref{bD}), i.e., common zeros $(c^2_*,\lambda^2_*,h^2_*,\Delta_*)$ of the beta functions, which correspond to possible (multi)critical points for the semimetal-superconductor transition. In the double epsilon expansion, the nature of the fixed points and their stability depend sensitively on the ratio $\epsilon/\epsilon_\tau$ (especially for disordered fixed points with $\Delta_*\neq 0$)~\cite{dorogovtsev1980,boyanovsky1982,lawrie1984}. Since we are interested in the limit $\epsilon\rightarrow 2$ and $\epsilon_\tau\rightarrow 1$, corresponding to 2+1 dimensions, we set $\epsilon=2\epsilon_\tau$ and expand to leading order in $\epsilon_\tau$.

\subsection{Fixed points}

\begin{figure}[t]
\includegraphics[width=0.8\columnwidth]{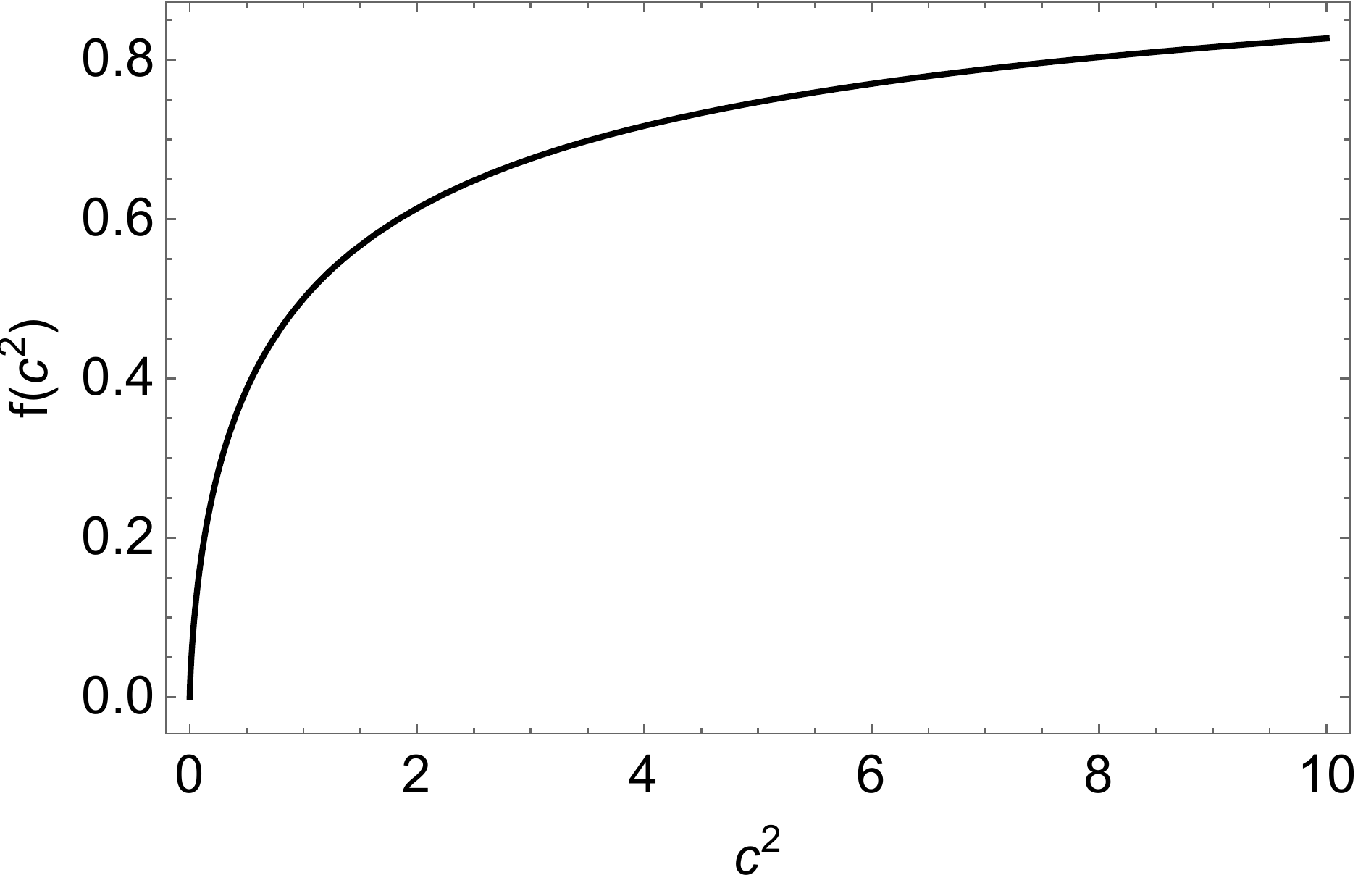}
\caption{Plot of $f(c^2)$ in Eq.~(\ref{f}), with $c^2=(c_f/c_b)^2$ the velocity ratio squared; $f(0)=0$, $f(1)=\frac{1}{2}$, and $f(\infty)=1$.}
\label{fig:f}
\end{figure}

First considering possible clean fixed points with $\Delta_*=0$, we find the Gaussian fixed point $(c_*^2,0,0,0)$ and $O(2)$ Wilson-Fisher fixed point $(c_*^2,\frac{\epsilon_\tau}{20},0,0)$, where $c_*^2$ is arbitrary since the velocity parameter flows under RG only in the presence of disorder or a nonzero Yukawa coupling [Eq.~(\ref{bc2})]. We also find a GNY fixed point for all $N$,
\begin{align}\label{CFP}
\left(1,\frac{2-N+\sqrt{N^2+76N+4}}{40(N+2)}\epsilon_\tau,\frac{\epsilon_\tau}{4(N+2)},0\right),
\end{align}
corresponding to the semimetal-superconductor QCP in the clean limit, and in agreement with earlier studies~\cite{rosenstein1993,roy2013,zerf2017,boyack2018}. Note that $\lambda_*^2>0$ for all $N\geq 1$. Since $f(1)=\frac{1}{2}$ (see Fig.~\ref{fig:f}), from Eq.~(\ref{z}) one finds $z=1$, and the clean QCP has emergent Lorentz invariance.

\begin{figure}[t]
\includegraphics[width=0.8\columnwidth]{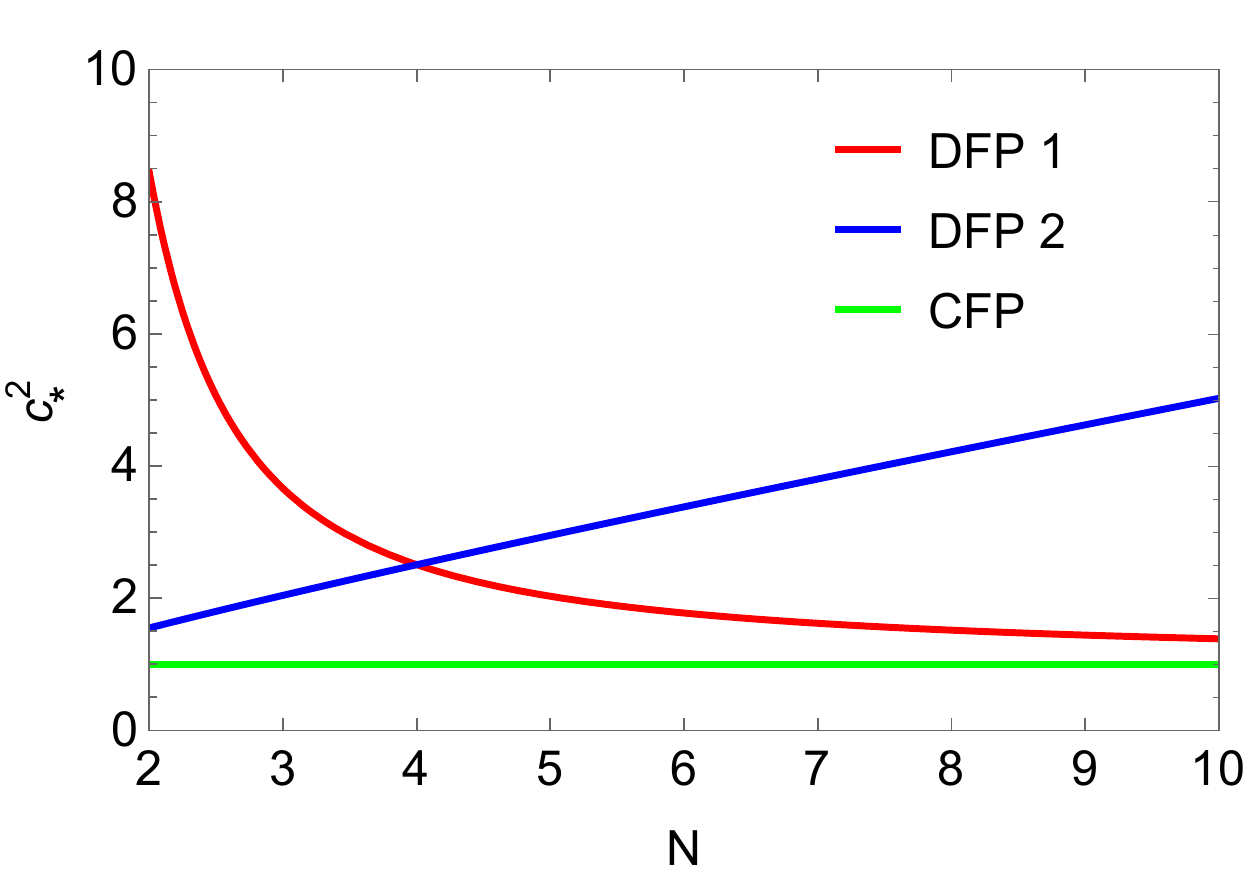}
\caption{Critical velocity parameters $c_*^2$ at the first disordered fixed point (DFP 1), the second disordered fixed point (DFP 2), and the clean fixed point (CFP, $c_*^2=1$), as a function of $N\geq 2$.}
\label{fig:c2vsN}
\end{figure}

We now look for possible disordered fixed points with $\Delta_*\neq 0$. Since at one-loop order $\beta_{h^2}$ depends on $h^2$ alone [Eq.~(\ref{bh2})], we can separately consider the cases with $h_*^2$ zero and nonzero. For $h_*^2=0$, we find the fixed point $(0,\frac{\epsilon_\tau}{2},0,\frac{3\epsilon_\tau}{4})$ for all $N$~\cite{NotezBC}, which corresponds to the disordered fixed point of the purely bosonic $O(2)$ model~\cite{dorogovtsev1980,boyanovsky1982,lawrie1984} and describes the superfluid-Mott glass transition in the presence of exact particle-hole symmetry~\cite{weichman2008}. For $h_*^2\neq 0$, as already mentioned one necessarily has $h_*^2=\epsilon_\tau/[4(N+2)]$ like at the clean fixed point (CFP) in Eq.~(\ref{CFP}), regardless of the values of $\lambda_*^2$ and $\Delta_*$. Solving for a common zero of $\beta_{\lambda^2}$ and $\beta_\Delta$, we find two nontrivial disordered fixed points (DFP),
\begin{align}
&\text{DFP 1: }\left(c_{*,\text{DFP1}}^2,\frac{\epsilon_\tau}{N+2},\frac{\epsilon_\tau}{4(N+2)},\frac{3\epsilon_\tau}{2(N+2)}\right),\label{DFP1}\\
&\text{DFP 2: }\left(c_{*,\text{DFP2}}^2,\frac{N\epsilon_\tau}{4(N+2)},\frac{\epsilon_\tau}{4(N+2)},\frac{(N-1)\epsilon_\tau}{2(N+2)}\right).\label{DFP2}
\end{align}
As they occur at finite Yukawa coupling, and thus involve strongly coupled bosonic \emph{and} fermionic critical fluctuations, we will term these fixed points \emph{fermionic} disordered fixed points. The critical couplings $\lambda_*^2$, $h_*^2$, and $\Delta_*$ are strictly positive, and thus physical, for all $N\geq 2$. Inserting (\ref{DFP1}) and (\ref{DFP2}) into $\beta_{c^2}$, one numerically finds that in both cases $\beta_{c^2}$ has a unique zero at a positive value of $c^2$ for all $N\geq 2$ (Fig.~\ref{fig:c2vsN}). For DFP 1, one can derive the lower bound $c_{*,\text{DFP1}}^2\geq N/(N-1)$, and $c_{*,\text{DFP1}}^2$ tends to one as $N$ increases. For DFP 2, $c_{*,\text{DFP2}}^2$ increases without bound as $N$ increases, and we have $c_{*,\text{DFP2}}^2\geq N/3$.

The cases $N=1$ and $N=4$ are special. As $N$ approaches one from above, DFP 2 merges with the clean fixed point, with $c_{*,\text{DFP2}}^2\rightarrow c_{*,\text{CFP}}^2=1$, while DFP 1 moves off to infinite coupling ($c_{*,\text{DFP1}}^2\rightarrow\infty$). As can be gleaned by looking at Eq.~(\ref{DFP1})-(\ref{DFP2}) and Fig.~\ref{fig:c2vsN}, as $N\rightarrow 4$ DFP 1 and DFP 2 also merge. In accordance with the general scenario governing the pairwise merging and annihilation of fixed points~\cite{kaplan2009}, and as will be elaborated upon below, in the presence of disorder we expect to find marginal scaling at the clean fixed point for $N=1$ and at the (unique) fermionic disordered fixed point for $N=4$.

\subsection{Linear stability analysis}
\label{sec:stability}

We now perform a linear stability analysis for the fixed points found in the previous section, within the critical hypersurface $r=0$. In the absence of disorder, as found previously~\cite{rosenstein1993,roy2013,zerf2017,boyack2018} the Gaussian and $O(2)$ Wilson-Fisher fixed points have at least one unstable direction, while the CFP is stable and describes the critical behavior at the transition. In the presence of disorder, both the Gaussian and $O(2)$ Wilson-Fisher fixed points acquire an additional unstable direction. At the CFP, the RG eigenvalue (defined as the negative of the slope of the ultraviolet beta functions) corresponding to disorder is
\begin{align}\label{y4CFP}
-\frac{2}{5}\left(\frac{\sqrt{N^2+76N+4}-N-8}{N+2}\right)\epsilon_\tau,
\end{align}
which is strictly negative for all $N\geq 2$. Thus disorder is perturbatively irrelevant at the CFP for all $N\geq 2$. For $N=1$, the eigenvalue (\ref{y4CFP}) vanishes and one has marginal scaling, as expected from the discussion at the end of the last section. Expanding the beta functions to quadratic order in the couplings near the CFP, we find that disorder is marginally relevant.

Turning now to the disordered fixed points, we find that the disordered $O(2)$ Wilson-Fisher fixed point is destabilized by a nonzero Yukawa coupling for all $N$. By contrast, the stability of DFP 1 and DFP 2 depends on $N$. For $N=2,3$, DFP 1 is stable while DFP 2 has one unstable direction; for $N=4$, DFP 1 and DFP 2 merge into a single fermionic disordered fixed point with marginal flow; for $N\geq 5$, DFP 1 and DFP 2 exchange their stability properties, i.e., DFP 2 is stable and DFP 1 has one unstable direction. As previously mentioned, for $N=1$ no finite-disorder fixed points remain.

\begin{figure}[t]
\includegraphics[width=0.9\columnwidth]{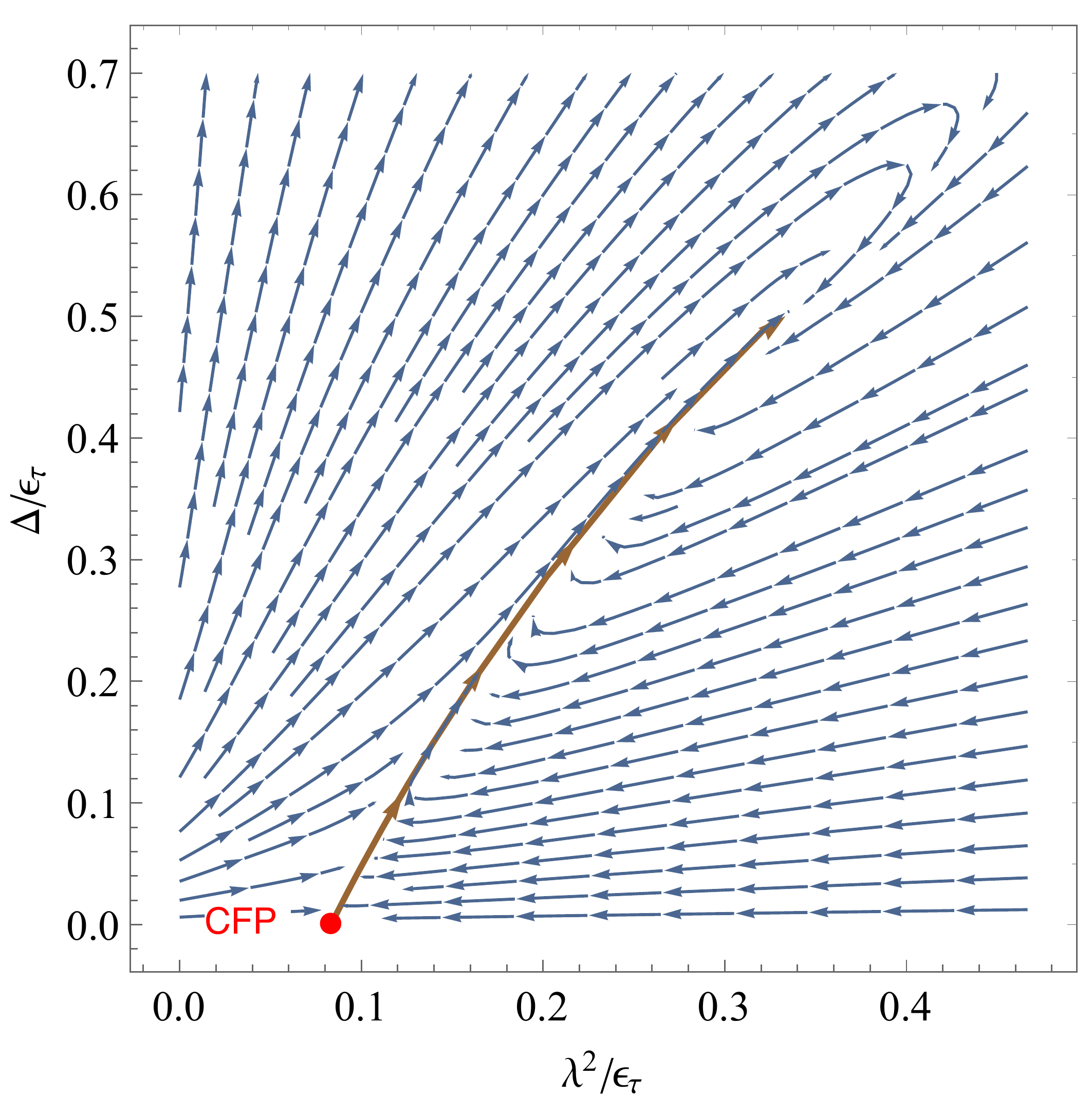}
\caption{RG flows for $N=1$, with marginal flow (brown line) away from the CFP.}
\label{fig:N=1}
\end{figure}

\subsection{RG flows}
\label{sec:RGflows}

Having investigated the linearized RG flow near the fixed points, we now analyze the full flow in the four-dimensional space of couplings, as given by the solution of the coupled differential equations (\ref{bc2})-(\ref{bD}). Since the beta function for the Yukawa coupling (\ref{bh2}) is independent of $c^2$, $\lambda^2$, and $\Delta$, the CFP, DFP 1, and DFP 2 share a common fixed-point value of $h^2_*=\epsilon_\tau/[4(N+2)]$. Furthermore, we find that the scaling field corresponding to the relative velocity parameter $c^2$ is irrelevant at each of those fixed points (except for $N=1$, which is discussed separately below). Therefore we will plot the projection of the RG flow in the $\lambda^2$-$\Delta$ plane at fixed $h^2=h_*^2$.

In Fig.~\ref{fig:N=1} we plot the projected RG flows for $N=1$. There is marginal flow away from the CFP, with nonzero projections along the $\lambda^2$, $\Delta$, and $c^2$ directions. The point $(\lambda,\Delta)=(\epsilon_\tau/3,\epsilon_\tau/2)$ towards which the marginal flow leads in Fig.~\ref{fig:N=1} is a remnant of DFP 1 [see Eq.~(\ref{DFP1})], but is not a fixed point as it is impossible to make $\beta_{c^2}$ vanish there for $N=1$. The marginal flow at the CFP implies the existence of a Landau pole that can be interpreted as a crossover temperature scale $T^*\sim\Lambda e^{-1/\alpha\Delta_0}$ above which scaling in the quantum critical fan is controlled by the CFP, where $\Lambda$ is a high-energy cutoff, $\Delta_0$ is a dimensionless measure of the bare disorder strength, and $\alpha$ is a numerical factor of order unity. Below $T^*$ the runaway flow suggests the existence of a new fixed point, not accessible at one-loop order, or a first-order transition.

\begin{figure}[t]
\includegraphics[width=0.9\columnwidth]{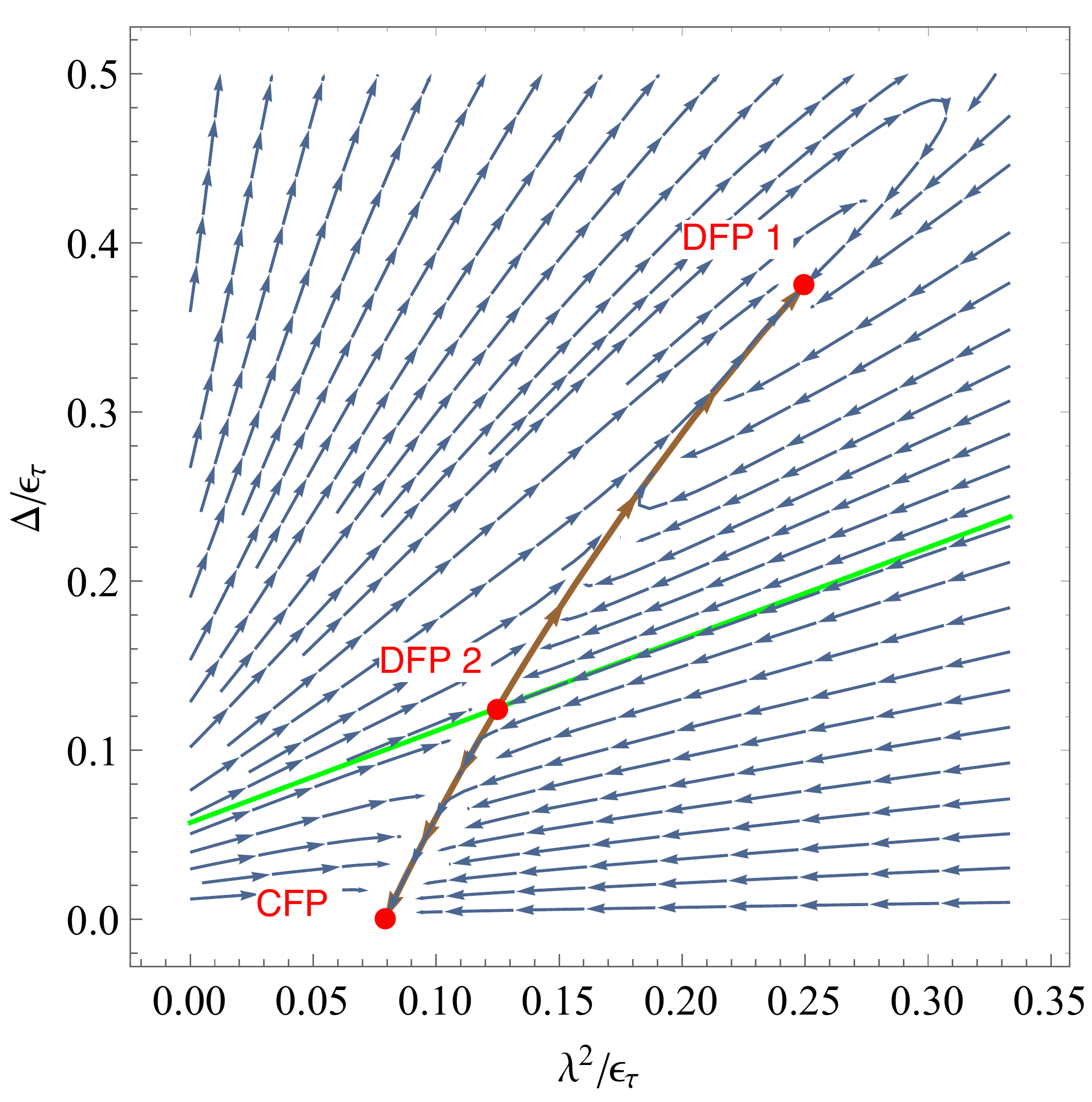}
\caption{RG flows for $N=2$, with separatrix (green line) controlled by DFP 2 between the CFP and DFP 1.}
\label{fig:N=2}
\end{figure}

\begin{figure}[t]
\includegraphics[width=0.9\columnwidth]{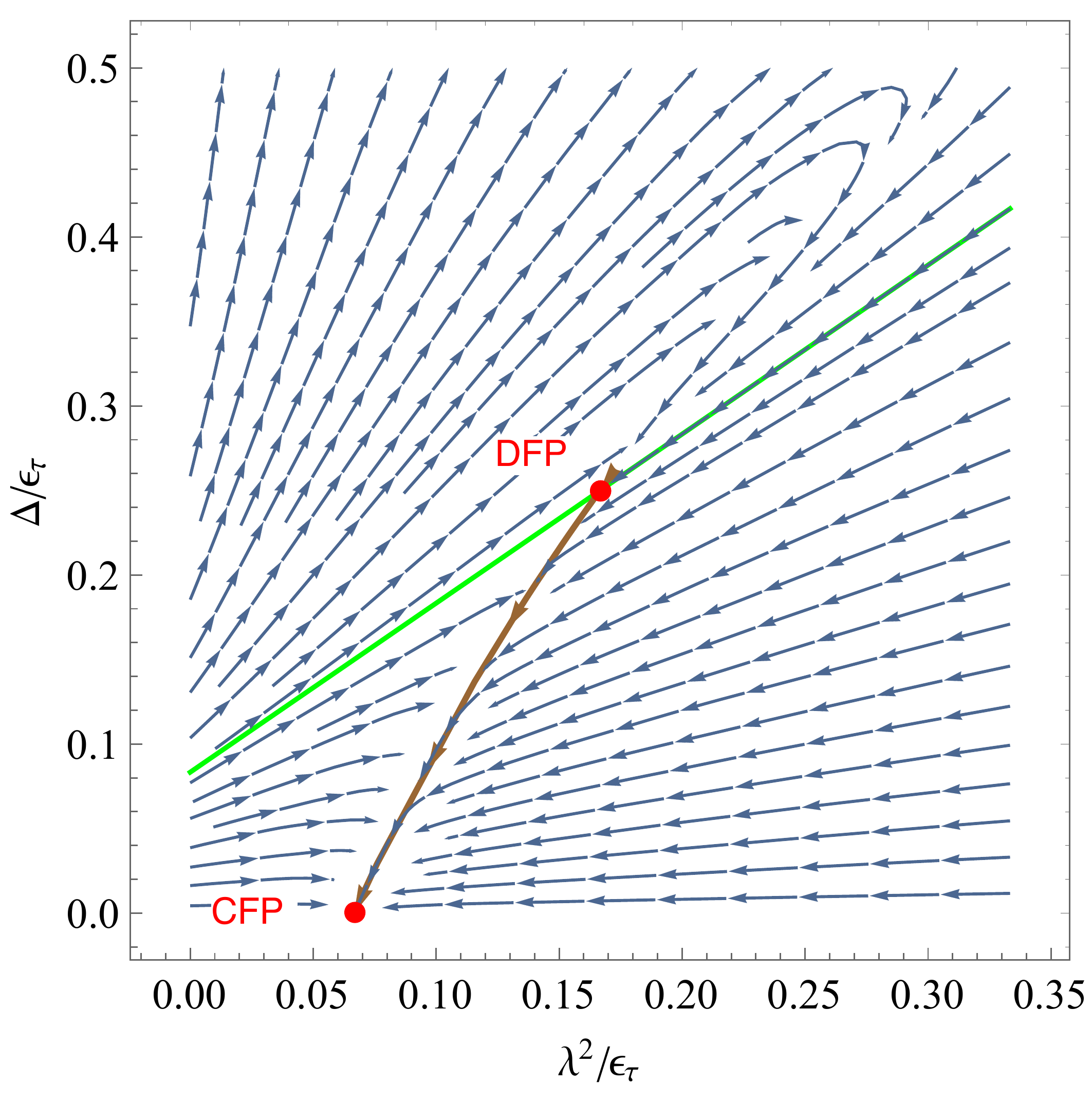}
\caption{RG flows for $N=4$: DFP 1 and DFP 2 merge into a single DFP with marginal flow towards the CFP.}
\label{fig:N=4}
\end{figure}

\begin{figure}[t]
\includegraphics[width=0.9\columnwidth]{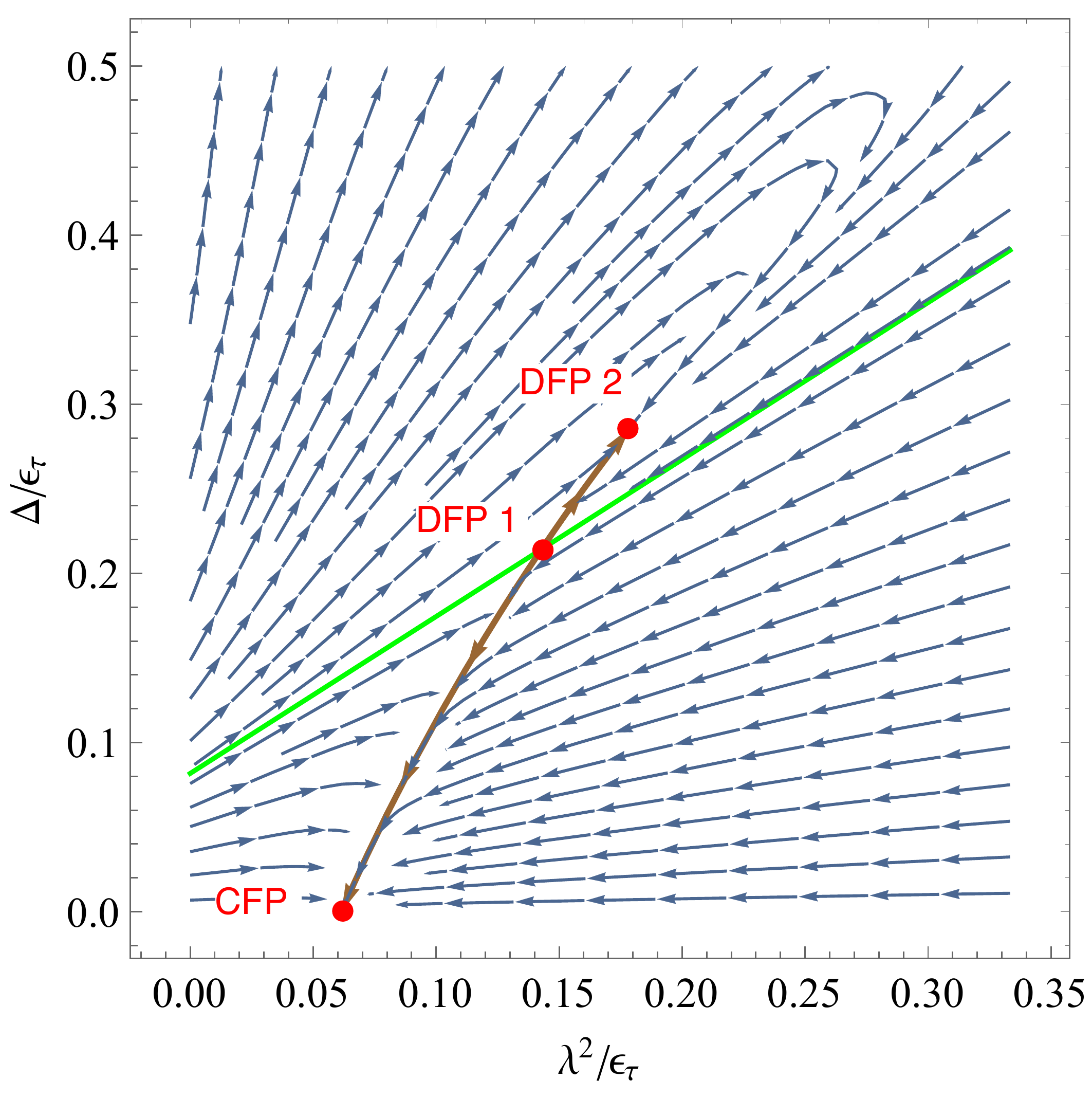}
\caption{RG flows for $N=5$.}
\label{fig:N=5}
\end{figure}

\begin{figure}[t]
\includegraphics[width=0.9\columnwidth]{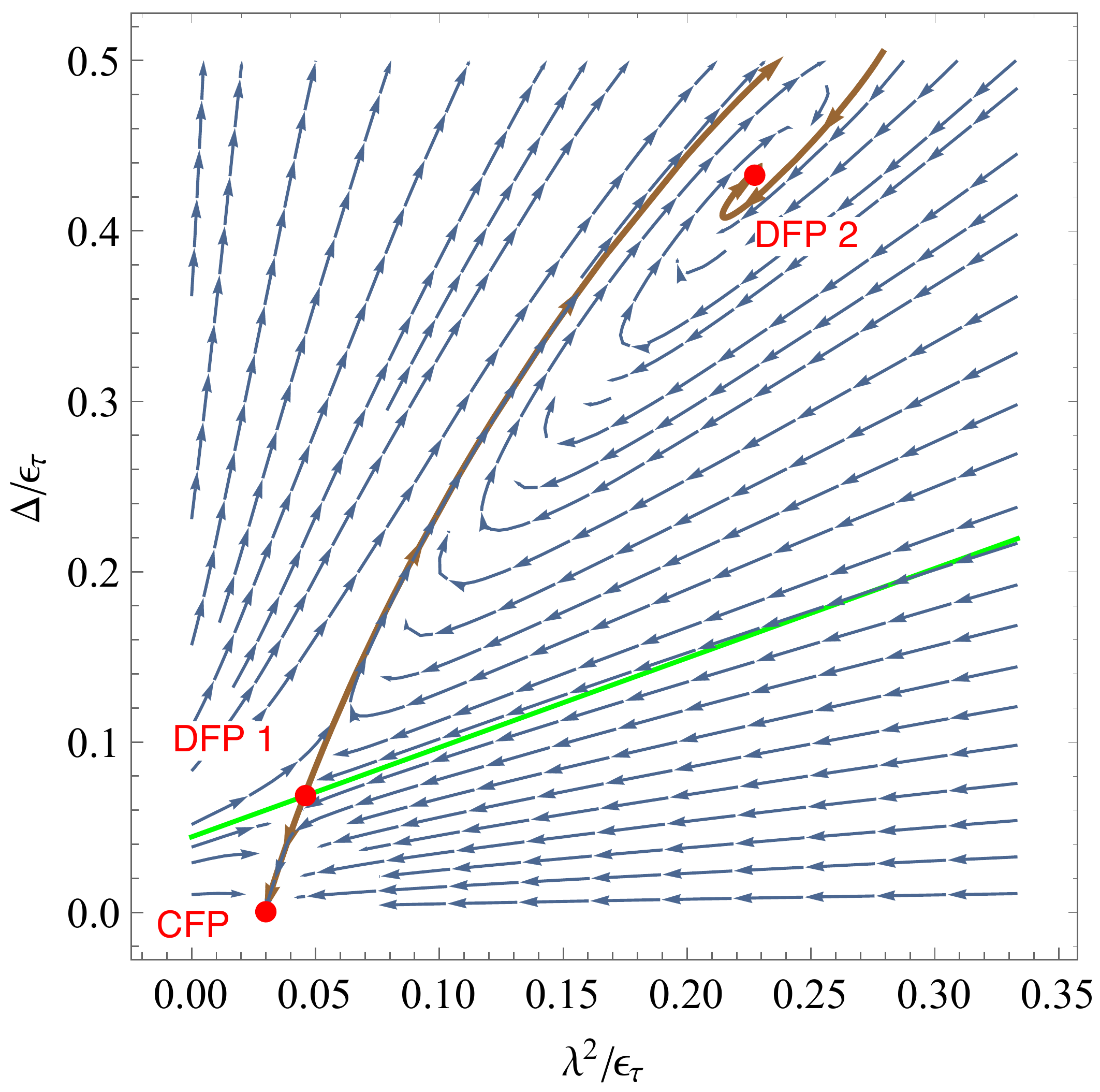}
\caption{RG flows for $N=20$; DFP 2 is a fixed point of stable-focus type for all $N\geq 7$.}
\label{fig:N=20}
\end{figure}

In Fig.~\ref{fig:N=2} we plot the flow diagram for $N=2$. As found in the linear stability analysis, the CFP and DFP 1 are stable fixed points while DFP 2 has one unstable direction, and controls a separatrix surface (appearing as a line in the $\lambda^2$-$\Delta$ plane) that separates the basins of attraction of the CFP and DFP 1. For $N=3$, the flow diagram is qualitatively similar but DFP 1 and DFP 2 approach each other; at $N=4$ they merge into a single DFP with marginal flow towards the CFP (Fig.~\ref{fig:N=4}).

For $N=5$ (Fig.~\ref{fig:N=5}) and $N=6$, the flow diagram is qualitatively similar as that for $N=2$ and $N=3$, but the stability properties of DFP 1 and DFP 2 are interchanged. DFP 1 now controls the separatrix, and DFP 2 is the stable fixed point. For $N\geq 7$, this state of affairs remains, but two irrelevant eigenvalues of the stability matrix acquire a nonzero imaginary part. Since the stability matrix is real, they are complex conjugates $\omega_\pm=\omega'\pm i\omega''$, but their real part $\omega'$ (defining $\omega_\pm$ to be the eigenvalues of the Jacobian matrix of the ultraviolet beta functions) remains positive, since they correspond to irrelevant directions. We obtain
\begin{align}\label{omegacmplx}
\omega_\pm=\frac{N+8\pm i\sqrt{3N(5N-32)}}{2(N+2)}\epsilon_\tau.
\end{align}
As a consequence of the nonzero imaginary part, RG trajectories spiral around DFP 2, and the latter becomes a fixed point of stable-focus type. Such fixed points have been found before in disordered $O(n)$ magnets~\cite{khmelnitskii1978,dorogovtsev1980}. As an illustrative example, we plot the RG flows for $N=20$ in Fig.~\ref{fig:N=20} (stable-focus behavior is obtained for all $N\geq 7$, but $\omega''$ is larger --- and thus the spiraling trajectories more easily seen --- for larger $N$.)

\section{Critical exponents and phase diagram}
\label{sec:critical}

From Eq.~(\ref{nuinv})-(\ref{gammaphi}) and the fixed point couplings (\ref{CFP}), (\ref{DFP1}), (\ref{DFP2}) we can now determine the critical exponents at the various fixed points (Table~\ref{table:exponents}), where $\eta_\psi$, $\eta_\phi$ denote the anomalous dimensions $\gamma_\psi$, $\gamma_\phi$ evaluated at the fixed point.

\begin{widetext}
\begin{center}
\begin{table}[h]
\centering
\begin{tabular}{ |c||c|c|c|c| } 
 \hline
Fixed point  & $\nu^{-1}$ & $z-1$ & $\eta_\psi$ & $\eta_\phi$ \\ 
  \hline
  \hline
CFP & $2-\left(\displaystyle\frac{4N+2+\sqrt{N^2+76N+4}}{5(N+2)}\right)\epsilon_\tau$ & 0 & $\displaystyle\frac{\epsilon_\tau}{N+2}$ & $\displaystyle\frac{N\epsilon_\tau}{N+2}$ \\
\hline
DFP 1 & $2-\left(\displaystyle\frac{N+5}{N+2}\right)\epsilon_\tau$ & $\displaystyle\frac{3+\left(\frac{1-c_*^2}{c_*^2}\right)N}{2(N+2)}\epsilon_\tau$ & $\displaystyle\frac{\epsilon_\tau}{N+2}$ & $\displaystyle\frac{N\epsilon_\tau}{N+2}$ \\
\hline
DFP 2 & $2-\left(\displaystyle\frac{2N+1}{N+2}\right)\epsilon_\tau$ & $\displaystyle\frac{\frac{N}{c_*^2}-1}{2(N+2)}\epsilon_\tau$ & $\displaystyle\frac{\epsilon_\tau}{N+2}$ & $\displaystyle\frac{N\epsilon_\tau}{N+2}$ \\
 \hline
\end{tabular}
\caption{Critical exponents at the CFP, DFP 1, and DFP 2.}
\label{table:exponents}
\end{table}
\end{center}
\end{widetext}

For $N=1$, the CFP becomes the supersymmetric fixed point with $\eta_\psi=\eta_\phi=\epsilon_\tau/3$~\cite{balents1998,ScottThomas,lee2007,grover2014,ponte2014,zerf2016,fei2016,zerf2017}. At the present one-loop order, the fermion/boson anomalous dimensions $\eta_\phi$ and $\eta_\psi$ only depend on the Yukawa coupling $h^2$, which is the same at each fixed point as observed earlier. This state of affairs will change at higher loop orders, and we expect the anomalous dimensions to differ at different fixed points in general.

\begin{figure}[t]
\includegraphics[width=0.8\columnwidth]{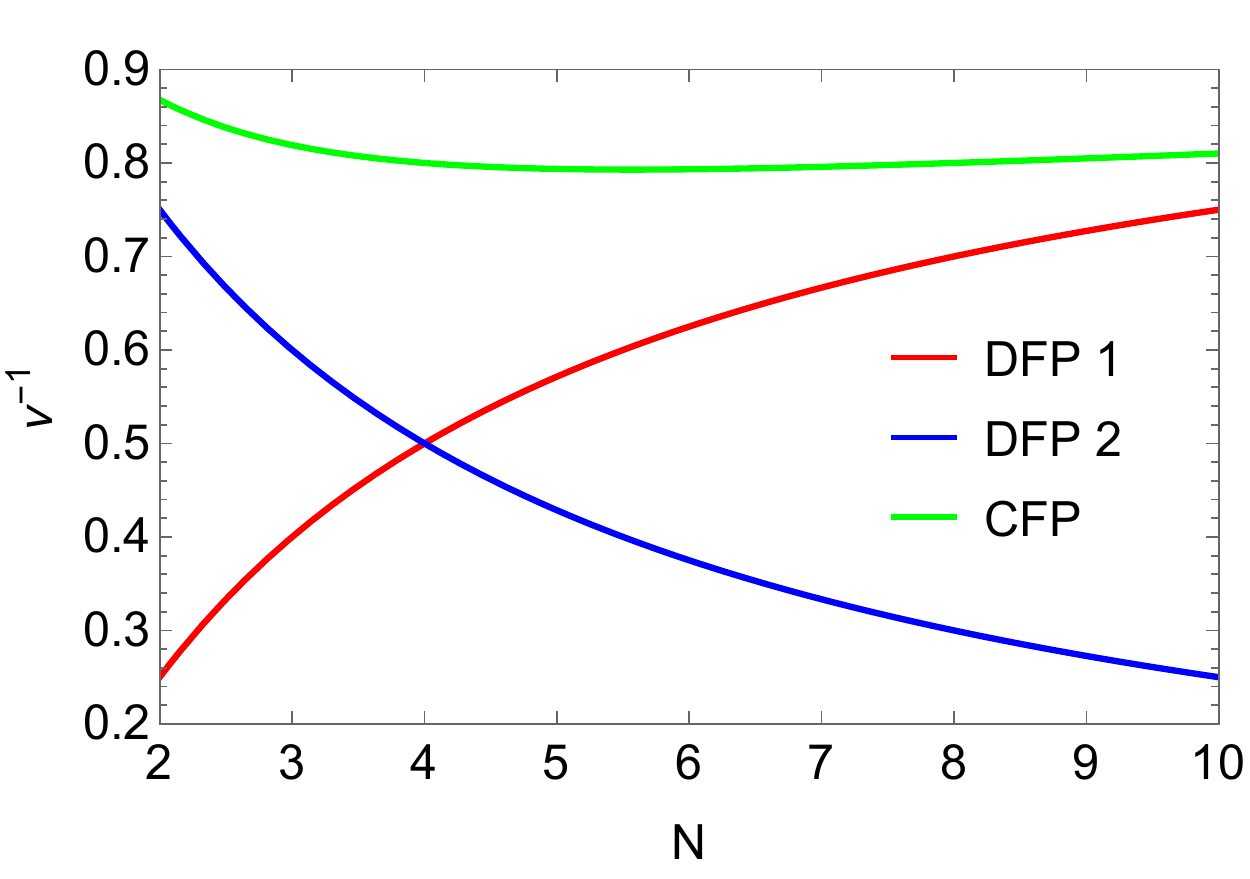}
\caption{Inverse correlation length exponent $\nu^{-1}$ for $\epsilon_\tau=1$, as a function of $N\geq 2$.}
\label{fig:nu}
\end{figure}

We plot the inverse correlation length exponent $\nu^{-1}$ extrapolated to $\epsilon_\tau=1$ as a function of $N\geq 2$ in Fig.~\ref{fig:nu}. In accordance with the linear stability analysis in Sec.~\ref{sec:stability}, the CFP obeys the Harris criterion~\cite{harris1974}, according to which clean critical behavior is stable against random-$T_c$ disorder if 
\begin{align}\label{harris}
\nu^{-1}<d/2,
\end{align}
where $d=2$ is the (physical) spatial dimension and $\nu^{-1}$ is the inverse correlation length exponent in the clean limit. At the CFP, $\nu^{-1}$ is strictly less than one for all $1<N<\infty$ and reaches one at both $N=1$ and $N=\infty$; thus for $N=1$ the CFP is Harris marginal, as found in Sec.~\ref{sec:RGflowanalysis}. Note that in the context of a perturbative RG analysis, it is more appropriate to use the Harris criterion in the form (\ref{harris}), rather than in the usual form $\nu>2/d$, as (\ref{harris}) simply expresses the condition of perturbative irrelevance of the disorder-induced interaction (\ref{SdisBos}), namely that its scaling dimension $2(d+\epsilon_\tau-\nu^{-1})$ be larger than the effective spacetime dimensionality $d+2\epsilon_\tau$ appropriate for this interaction. However, this makes clear the fact that the Harris criterion is one of perturbative stability, and does not preclude the existence of disordered critical points occurring past a certain finite critical disorder strength, as found here. At the DFP 1 (DFP 2), $\nu^{-1}$ increases (decreases) monotonically as $N$ increases, asymptotically reaching $1$ ($0$) at $N=\infty$. Thus at all fixed points $\nu^{-1}\leq 1$, in agreement with the Chayes inequality $\nu^{-1}\leq d/2$ for critical points in disordered systems~\cite{chayes1986}.

\begin{figure}[t]
\includegraphics[width=0.8\columnwidth]{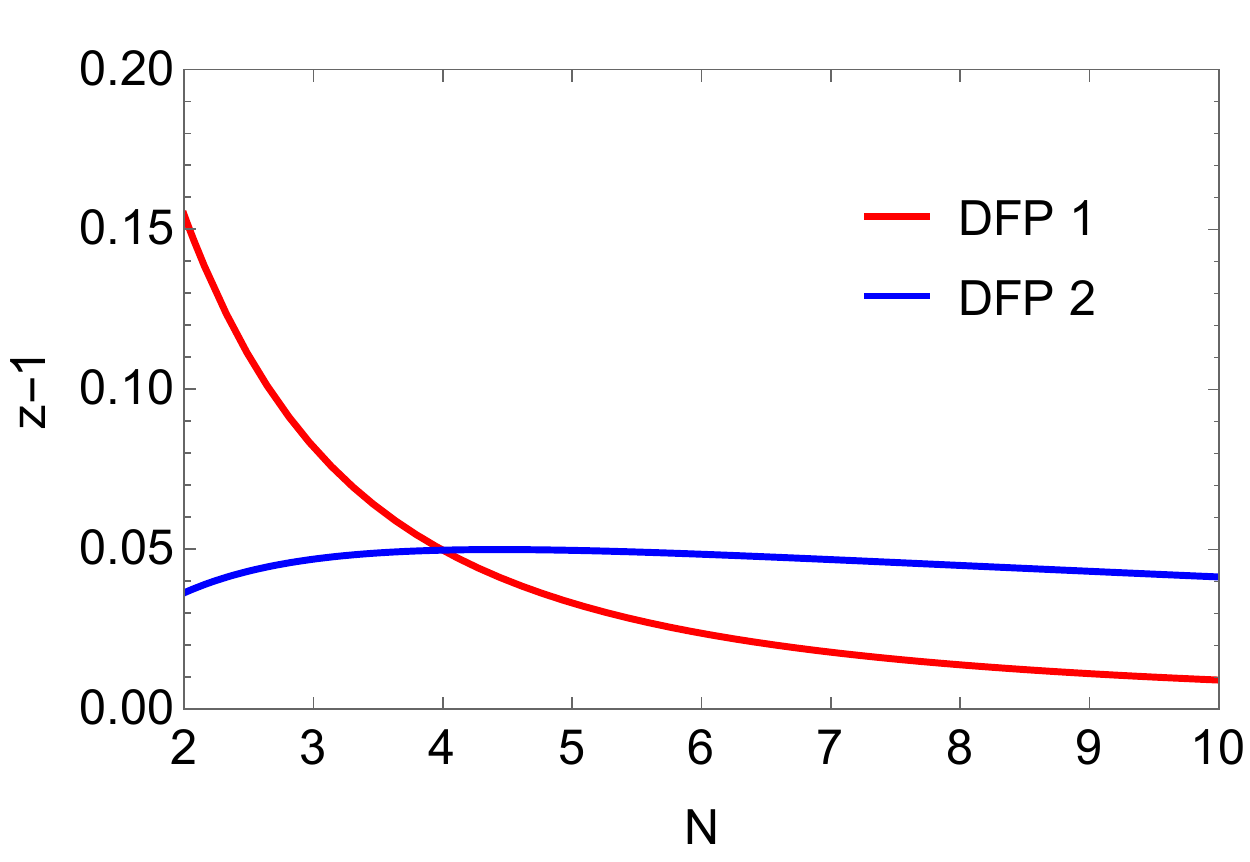}
\caption{Correction $z-1$ to the dynamic critical exponent for $\epsilon_\tau=1$ at the two disordered fixed points, as a function of $N$.}
\label{fig:z}
\end{figure}

We also plot the deviation of the dynamic critical exponent $z$ from unity at DFP 1 and DFP 2 in Fig.~\ref{fig:z}, as a function of $N\geq 2$, and extrapolated to $\epsilon_\tau=1$ (or equivalently, in units of $\epsilon_\tau$). The dynamic critical exponent depends on the fixed-point value of the relative velocity parameter $c^2_*$, itself plotted in Fig.~\ref{fig:c2vsN}.

Finally, by contrast with standard RG fixed points of source/sink type where RG trajectories approach the fixed point monotonically, fixed points of stable-focus type, such as the DFP 2 for $N\geq 7$, are known to lead to oscillatory corrections to scaling laws~\cite{khmelnitskii1978}. In particular, the uniform, static order parameter susceptibility $\chi$, which obeys the usual scaling law $\chi\sim |r|^{-\gamma}$ with $\gamma$ the susceptibility exponent, develops corrections of the form
\begin{align}\label{oscillatory}
\chi\sim|r|^{-\gamma}\left[1+C\left|\frac{r}{r_0}\right|^{\nu\omega'}\cos\left(\nu\omega''\ln\left|\frac{r}{r_0}\right|+\phi\right)+\ldots\right],
\end{align}
where $r_0$, $C$, and $\phi$ are nonuniversal constants that depend on the initial distance to the fixed point within the critical hypersurface $r=0$, but the exponents $\omega'$ and $\omega''$, given in Eq.~(\ref{omegacmplx}) and plotted in Fig.~\ref{fig:corrscaling}, are universal properties of the fixed point. [See Appendix~\ref{app:oscillatory} for a derivation of Eq.~(\ref{oscillatory}).]

\begin{figure}[t]
\includegraphics[width=0.8\columnwidth]{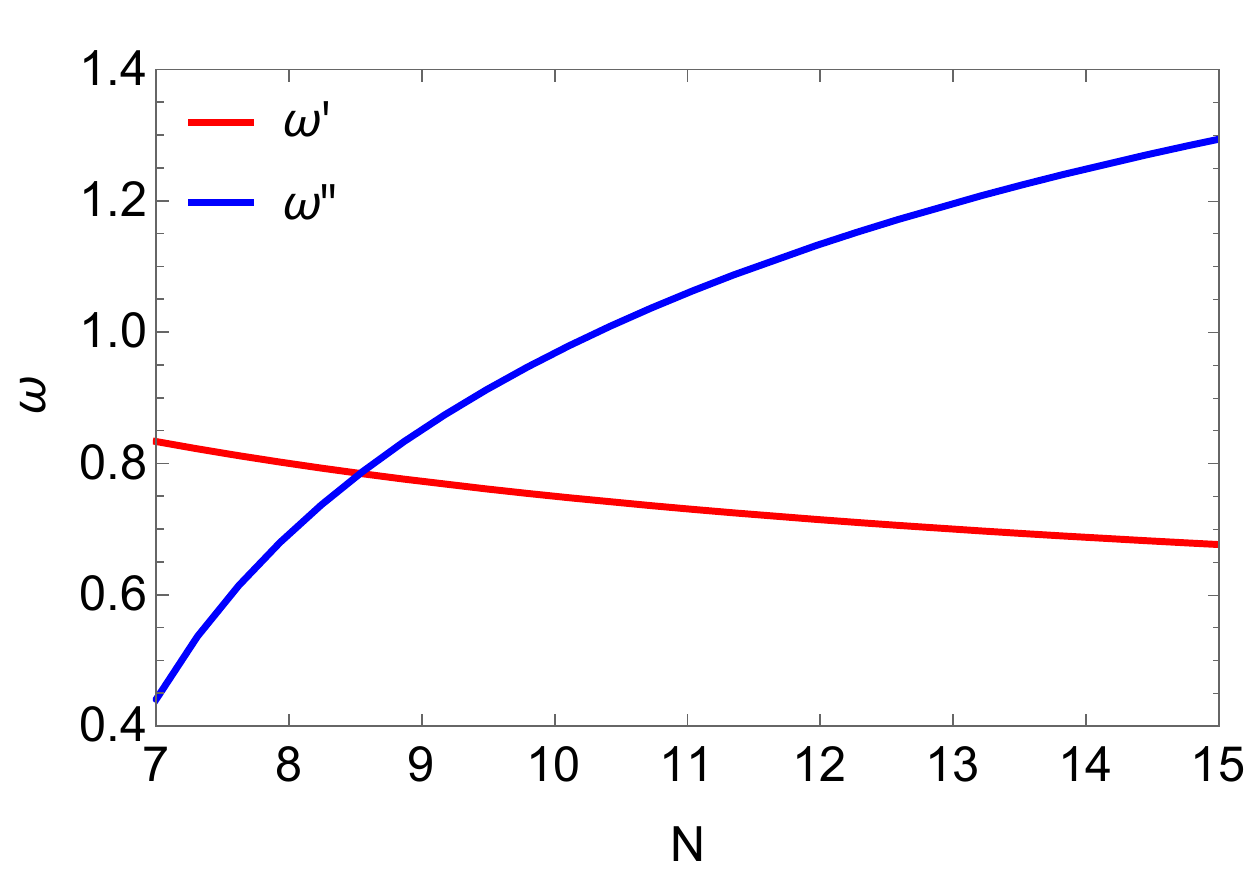}
\caption{Exponents $\omega'$ and $\omega''$ appearing in oscillatory corrections to scaling at DFP 2 for $N\geq 7$, for $\epsilon_\tau=1$.}
\label{fig:corrscaling}
\end{figure}

\begin{figure}[t]
\includegraphics[width=0.8\columnwidth]{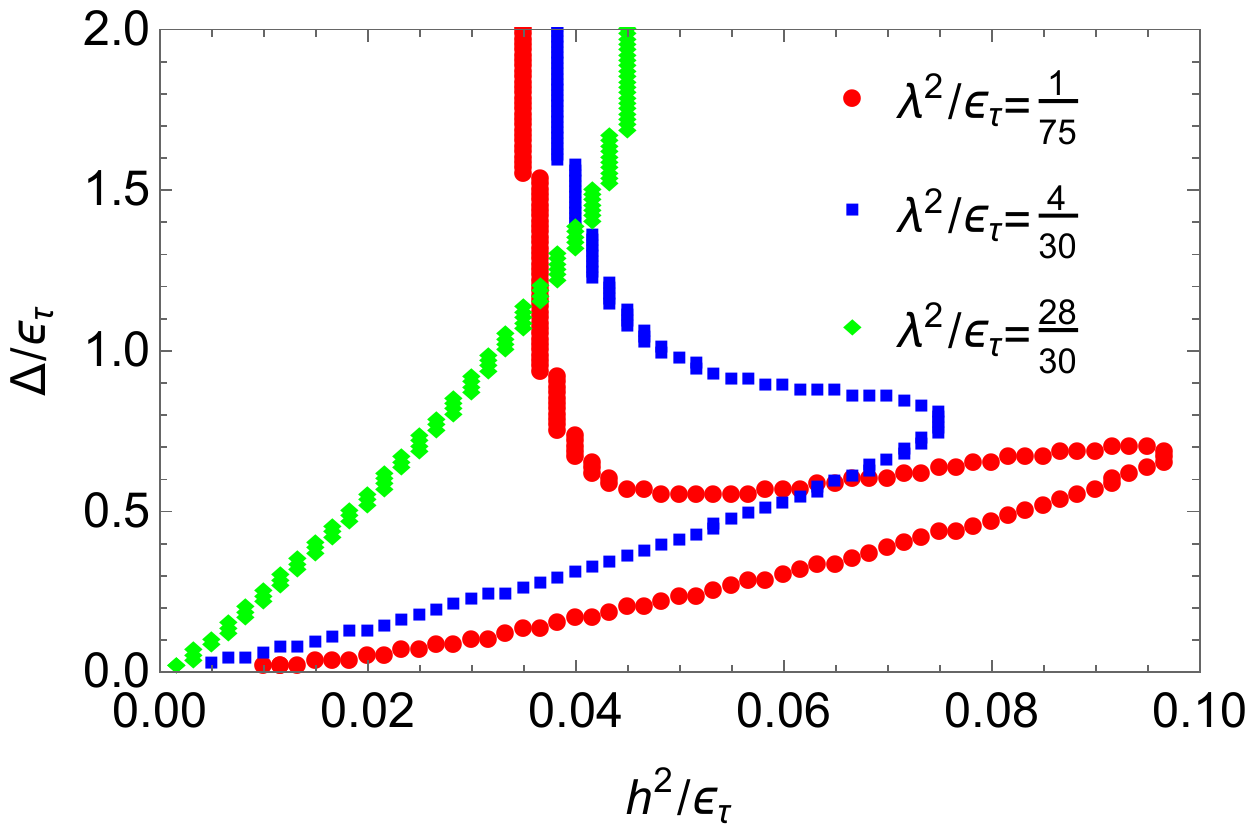}
\caption{Cuts of the separatrix surface at constant $\lambda^2$ for $N=8$.}
\label{fig:separatrix}
\end{figure}

The separatrix surface for $N\geq 2$ mentioned in Sec.~\ref{sec:RGflows} has interesting nonmonotonicity properties. As the direction corresponding to the relative velocity parameter $c^2$ is always irrelevant at the CFP, DFP 1, and DFP 2 for $N\geq 2$, it is sufficient to consider the separatrix as a 2D surface in the 3D reduced parameter space $(\lambda^2,h^2,\Delta)$. In Fig.~\ref{fig:separatrix} we plot three cuts through this surface at constant $\lambda^2$ that are representative of the qualitative behavior we have observed numerically for all $N\geq 2$, and which can be summarized as follows. Let $\Delta=g_{\lambda^2}(h^2)$ be an equation describing the separatrix curve in the $h^2$-$\Delta$ plane for a given $\lambda^2$. Then there always exists an interval $[h_1^2,h_2^2]$, dependent on $\lambda^2$, and a value $\lambda_1^2$ such that for $\lambda^2<\lambda_1^2$, the function $g_{\lambda^2}(h^2)$ is double valued. Conversely, consider describing the same separatrix curve by the equation $h^2=g^{-1}_{\lambda^2}(\Delta)$ where $g^{-1}$ is the inverse function. Then likewise there always exists an interval $[\Delta_1,\Delta_2]$, dependent on $\lambda^2$, and a value $\lambda_2^2<\lambda_1^2$ such that for $\lambda^2<\lambda_2^2$ the function $g^{-1}_{\lambda^2}(\Delta)$ is double valued. This double-valued/nonmonotonic behavior of the separatrix surface has potential consequences for the phase diagram of the system as will be discussed below.

By following the RG trajectories from a set of initial conditions for the coupling constants $(c^2,\lambda^2,h^2,\Delta)$ one can deduce the following implications for the phase diagram of the system. The $N=1$ case has already been discussed previously: the one-loop analysis does not allow one to determine the ultimate fate of the quantum critical point, which can either fall in a new universality class or become a first-order transition. For $N\geq 2$, consider as tuning variables the critical tuning parameter for the transition, $r$, and the disorder strength $\Delta$, assuming that $\lambda^2$ and $h^2$ are held fixed. For $\Delta=0$ the transition is between a clean Dirac semimetal and a superconductor, and is in the universality class of the CFP. For sufficiently small nonzero $\Delta$, the initial conditions in parameter space remain in the basin of attraction of the CFP and the universality class of the transition is still controlled by the latter. While irrelevant at the critical point in the double epsilon expansion, chemical potential disorder --- which led to the disorder-induced four-fermion interaction in Eq.~(\ref{SdisMu}) --- is known to generate a nonzero density of states at (2+1)D Dirac points in the absence of electron-electron interactions, producing diffusive metallic behavior~\cite{fisher1985,fradkin1986b,ludwig1994}. In other words, Eq.~(\ref{SdisMu}) can be thought of as a dangerously irrelevant interaction. Note that we considered sufficiently smooth disorder, such that there is no backscattering between different Dirac points and thus no localization effects. As a result, for $\Delta >0$ the transition is really from a diffusive metal to a superconductor. Rare-region effects will likely lead to the formation of quantum Griffiths phases on both sides of the transition~\cite{vojta2013}, characterized by essential Griffiths-McCoy singularities, but are expected to produce exponentially small corrections to thermodynamic observables at the critical point~\cite{vojta2005}.

\begin{figure}[t]
\includegraphics[width=\columnwidth]{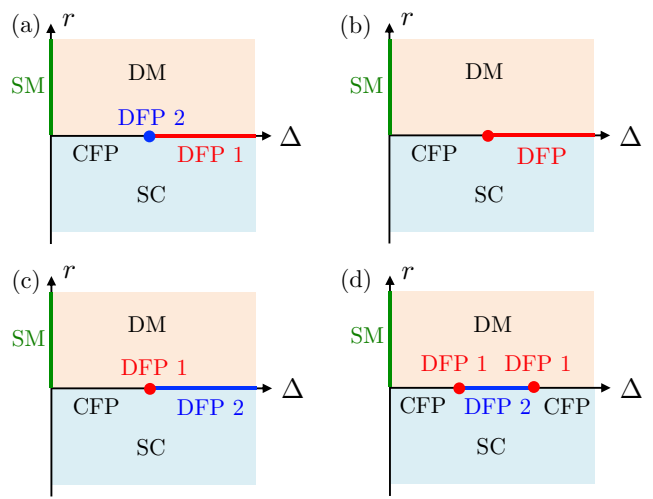}
\caption{Schematic phase diagrams in the plane of tuning parameter $r$ and disorder strength $\Delta$ for $N\geq 2$. SM: Dirac semimetal; DM: diffusive metal; SC: superconductor. For sufficiently small initial values of $\lambda^2$ and $h^2$, the universality class of the transition changes beyond a critical disorder strength from that of the CFP to that of one of the two disordered fixed points: (a) $N=2$ and $N=3$; (b) $N=4$; (c) $N\geq 5$. For sufficiently large $\lambda^2$ and/or $h^2$, beyond a second critical disorder strength there is a reentrant critical regime controlled by the CFP [plotted in (d) for $N\geq 5$, but an analogous effect occurs for $2\leq N\leq 4$].}
\label{fig:phasediagram}
\end{figure}

As $\Delta$ increases, it eventually crosses the separatrix surface at a certain critical value $\Delta_{c,1}$, and for $\Delta>\Delta_{c,1}$ enters the basin of attraction of a disordered fixed point. Thus for $N=2$ and $N=3$, the universality class of the transition is controlled by the CFP for $\Delta<\Delta_{c,1}$, by DFP 2 for $\Delta=\Delta_{c,1}$, which is a multicritical point, and by DFP 1 for $\Delta>\Delta_{c,1}$ [see Fig.~\ref{fig:phasediagram}(a)]. For $N=4$, for $\Delta>\Delta_{c,1}$ the RG trajectories flow back to the (unique) DFP, such that the universality class of the transition is controlled by the DFP for $\Delta\geq\Delta_{c,1}$ [Fig.~\ref{fig:phasediagram}(b)]. For $N\geq 5$, the scenario is the same as for $N=2$ and $N=3$ but the roles of DFP 1 and DFP 2 are exchanged, with DFP 1 acting as multicritical point at $\Delta=\Delta_{c,1}$ and DFP 2 controlling the critical behavior for $\Delta>\Delta_{c,1}$ [Fig.~\ref{fig:phasediagram}(c)].

As mentioned earlier and illustrated in Fig.~\ref{fig:separatrix}, for sufficiently small $\lambda^2$ there is always an interval of values of $h^2$ for which the separatrix curve is a double-valued function of $h^2$. As a result, if the initial value of $h^2$ is contained in this interval, as the disorder strength $\Delta$ increases from zero the universality class of the transition will be first controlled by the CFP, then by one of the disordered fixed points (depending on the value of $N$), and then again by the CFP [Fig.~\ref{fig:phasediagram}(d)]. However, this counterintuitive behavior may be an artefact of the one-loop approximation.

\section{Conclusion}
\label{sec:conclusion}

In conclusion, we have studied the critical properties of the semimetal-superconductor quantum phase transition in a model of 2D Dirac semimetal with $N$ flavors of two-component Dirac fermions, in the presence of quenched disorder assumed to be uncorrelated, but sufficiently smooth so as to make the probability of scattering between different Dirac cones negligible. Our one-loop analysis demonstrated the possibility of a general scenario for critical phenomena in disordered systems, to our knowledge not explicitly discussed in the literature so far: a clean critical point may be stable against disorder according to the Harris criterion, but yet may be replaced by a finite-disorder critical point beyond a certain finite, critical disorder strength. In the model studied here such finite-disorder critical points were characterized by finite fixed-point values of both the boson-boson and fermion-boson couplings, and thus were dubbed disordered fermionic QCPs. Other notable features of the disordered critical points found included a noninteger dynamic critical exponent $z>1$, as well as oscillatory corrections to scaling for sufficiently large $N$.

Possible applications of our results include the semimetal-superconductor quantum phase transition in graphene ($N=4$) and on the surface of a 3D topological insulator ($N=1$); the experimental results reported in Ref.~\cite{zhao2015b} are encouraging in regards to the latter, although one would need to additionally tune the chemical potential to the Dirac point and reach the quantum critical regime by the application of a nonthermal tuning parameter such as pressure. With those caveats in mind, we also note that the surface of 3D topological crystalline insulators~\cite{fu2011,hsieh2012} such as SnTe~\cite{tanaka2012}, Pb$_{1-x}$Sn$_x$Se~\cite{dziawa2012}, and Pb$_{1-x}$Sn$_x$Te~\cite{xu2012} supports $N=4$ two-component Dirac cones, as in graphene, and that superconductivity has been observed in In-doped SnTe~\cite{sasaki2012,sato2013}, though presumably of bulk origin. Larger values of $N$ may be accessible in systems of ultracold large-spin alkaline-earth fermions~\cite{desalvo2010} loaded into optical honeycomb lattices, such as those studied theoretically in Ref.~\cite{zhou2016}, but with interactions tuned to be attractive. Alternatively, our results may be relevant for the Kekul\'e valence-bond-solid transition of repulsively interacting fermions on the honeycomb lattice, but the interplay of disorder with the $C_3$ point group symmetry, which is broken by the Kekul\'e order parameter, should be first investigated carefully. Besides the effect of disorder on the Kekul\'e transition, our approach can also be applied to other fermionic QCPs described by GNY-type theories, on which we will report in future publications.

To further elucidate the critical behavior at $N=1$ in the present model, perturbative calculations at two-loop order would be necessary. The conformal bootstrap~\cite{bobev2015}, perturbative RG studies of the clean chiral XY GNY model at four-loop order~\cite{zerf2017}, as well as quantum Monte Carlo simulations~\cite{li2017b} suggest that $\nu^{-1}$ is slightly above one at the CFP for $N=1$, implying via the Harris criterion that disorder is in fact relevant (as opposed to marginally relevant as found at one-loop order) at the CFP. (Interestingly, for $N=4$ quantum Monte Carlo simulations of the Kekul\'e transition in graphene~\cite{li2017} and naive extrapolation of the four-loop GNY $\epsilon$-expansion results~\cite{zerf2017} predict $\nu^{-1}>1$ at the CFP, while Pad\'e extrapolation of the latter results~\cite{zerf2017} as well as functional RG studies of the Kekul\'e transition~\cite{classen2017} predict $\nu^{-1}<1$ in the clean limit, in agreement with our one-loop result.) Beyond perturbative RG, it would be interesting to try to apply strong-disorder RG methods~\cite{dasgupta1980,bhatt1982,fisher1994} to this problem, as done recently for the 2D bosonic superfluid-Mott insulator transition~\cite{iyer2012}, or to incorporate the effect of quenched disorder in the sign-problem-free quantum Monte Carlo simulations of Ref.~\cite{li2017b}, as done previously for the disordered attractive Hubbard model~\cite{scalettar1999}.

\acknowledgments

We thank I. Affleck, F. Marsiglio, R. Nandkishore, A. Penin, A. Thomson, and A. Vishwanath for helpful correspondence. HY was supported by Alberta Innovates Technology Futures (AITF). JM was supported by NSERC grant \#RGPIN-2014-4608, the CRC Program, CIFAR, and the University of Alberta.

\appendix

\numberwithin{equation}{section}
\numberwithin{figure}{section}

\section{Relation between two-component and four-component formulations}
\label{app:spinors}

In this section we prove the equivalence between the two-component formulation of the chiral XY GNY model, used here and in Ref.~\cite{zerf2016}, and its four-component formulation, used in Ref.~\cite{roy2013,zerf2017}. We are only concerned with the fermion part of the Lagrangian, and will set $c_f=1$ for simplicity, without loss of generality. Consider an even number $N=2N_f$ of flavors of two-component Dirac fermions $\psi_\alpha$, $\alpha=1,\ldots,N$. Combining those into $N_f$ four-component Dirac spinors,
\begin{align}
\Psi_\alpha=\left(\begin{array}{c}
\psi_\alpha \\
i\psi_{\alpha+N_f}
\end{array}\right),\hspace{5mm}
\alpha=1,\ldots,N_f,
\end{align}
the fermion Lagrangian can be written as
\begin{align}\label{LMajoranaMass}
\mathcal{L}_f=\sum_{\alpha=1}^{N_f}\bar{\Psi}_\alpha\slashed{\partial}\Psi_\alpha+h\left(\phi^*\sum_{\alpha=1}^{N_f}
\Psi_\alpha^Ti\Gamma_2\Psi_\alpha+\text{H.c.}\right),
\end{align}
where $\bar{\Psi}_\alpha=\Psi_\alpha^\dag\Gamma_0$, $\slashed{\partial}=\Gamma_\mu\partial_\mu$, and we define the $4\times 4$ gamma matrices
\begin{align}\label{4x4Gamma}
\Gamma_\mu=\left(\begin{array}{cc}
\gamma_\mu & 0 \\
0 & -\gamma_\mu
\end{array}\right),\hspace{5mm}\mu=0,1,2.
\end{align}
One can easily check that the Lagrangian of Sec.~\ref{sec:model} is reproduced by a suitable choice of $2\times 2$ gamma matrices, such as $\gamma_0=\sigma_3$, $\gamma_1=\sigma_1$, and $\gamma_2=\sigma_2$. One can further define the two Hermitian matrices
\begin{align}
\Gamma_3=\left(\begin{array}{cc}
0 & -i \\
i & 0
\end{array}\right),\hspace{5mm}
\Gamma_5=\Gamma_0\Gamma_1\Gamma_2\Gamma_3=\left(\begin{array}{cc}
0 & 1 \\
1 & 0
\end{array}\right),
\end{align}
which square to the identity and anticommute with the gamma matrices (\ref{4x4Gamma}). Defining the charge conjugation matrix $C=i\Gamma_2$, we now perform a change of variables to a new set of $N_f$ four-component spinors $\chi_\alpha$~\cite{kleinert1998},
\begin{align}\label{Psia}
\Psi_\alpha=P_-\chi_\alpha+P_+C\bar{\chi}_\alpha^T,
\end{align}
where $P_\pm=\frac{1}{2}(1\pm\Gamma_5)$ are projectors obeying $P_\pm^2=P_\pm$ and $P_+P_-=P_-P_+=0$. Using the properties $C\Gamma_\mu C^{-1}=-\Gamma_\mu^T$ and $P_\pm\Gamma_\mu=\Gamma_\mu P_\mp$, $\mu=0,1,2$, the conjugate spinor is given by
\begin{align}\label{barPsia}
\bar{\Psi}_\alpha=\bar{\chi}_\alpha P_+ + \chi_\alpha^TCP_-.
\end{align}
Inserting Eq.~(\ref{Psia})-(\ref{barPsia}) into the Lagrangian (\ref{LMajoranaMass}), and using the properties $CP_\pm C^{-1}=P_\mp$, $P_\pm^T=P_\pm$, and $C^T=C^{-1}=C^\dag=-C$, we find
\begin{align}
\mathcal{L}_f=\sum_{\alpha=1}^{N_f}\bar{\chi}_\alpha\slashed{\partial}\chi_\alpha+2h\sum_{\alpha=1}^{N_f}
\bar{\chi}_\alpha(\phi_1+i\phi_2\Gamma_5)\chi_\alpha,
\end{align}
where $\phi=\phi_1+i\phi_2$, which is the form of the chiral XY GNY model given in Ref.~\cite{roy2013,zerf2017}. In graphene $N_f=2$, thus for us $N=2N_f=4$.

\section{Calculation of the renormalization constants at one-loop order}
\label{app:Z}

In this Appendix we calculate contributions to the divergent part of the one-loop 1PI effective action, $\Gamma_\text{div}$, that correspond to the Feynman diagrams in Fig.~\ref{fig:diagrams}. Demanding that the full renormalized 1PI effective action (including the counterterms) remains finite allows us to extract the one-loop contributions to the renormalization constants $\delta Z_i$, $i=1,\ldots,7,r$. At one-loop order there is no diagram consistent with the Feynman rules in Fig.~\ref{fig:feynrules} that can renormalize the Yukawa vertex; thus $\delta Z_6=0$ at this order.

\subsection{Boson two-point function}

The diagrams are given in Fig.~\ref{fig:diagrams}(a,b,c). Fig.~\ref{fig:diagrams}(a) and (c) are tadpole diagrams which contribute to the boson mass renormalization constant $Z_r$, thus in those diagrams one must use a massive boson propagator,
\begin{align}
D_{ab}(p)=\frac{\delta_{ab}}{c^2p_0^2+\b{p}^2+r\mu^2}.
\end{align}

For Fig.~\ref{fig:diagrams}(a), we obtain
\begin{align}
\delta\Gamma_\text{div}^\text{(a)}&=4\lambda^2\int\frac{d^{\epsilon_\tau}p_0}{(2\pi)^{\epsilon_\tau}}\int\frac{d^d\b{p}}{(2\pi)^d}\frac{1}{c^2p_0^2+\b{p}^2+r\mu^2}\nn\\
&\hspace{5mm}\times\sum_a\int d^d\b{x}\int d^{\epsilon_\tau}\tau\,|\phi_a|^2.
\end{align}
Here and in the rest of this Appendix momentum integrals are evaluated in the limit $\epsilon,\epsilon_\tau\rightarrow 0$ and discarding all finite terms. We obtain
\begin{align}
\int\frac{d^{\epsilon_\tau}p_0}{(2\pi)^{\epsilon_\tau}}\int\frac{d^d\b{p}}{(2\pi)^d}\frac{1}{c^2p_0^2+\b{p}^2+r\mu^2}=-\frac{r\mu^2}{8\pi^2(\epsilon-\epsilon_\tau)},
\end{align}
thus
\begin{align}
\delta Z_r^\text{(a)}=\frac{\lambda^2}{2\pi^2(\epsilon-\epsilon_\tau)}.
\end{align}

For Fig.~\ref{fig:diagrams}(c), ignoring a term which vanishes in the replica limit we have
\begin{align}
\delta\Gamma_\text{div}^\text{(c)}&=-\Delta\sum_a\int\frac{d^Dk}{(2\pi)^D}|\phi_a(k)|^2\nn\\
&\hspace{5mm}\times\int\frac{d^d\b{p}}{(2\pi)^d}\frac{1}{c^2k_0^2+\b{p}^2+r\mu^2},
\end{align}
where $d^Dk=d^{\epsilon_\tau}k_0\,d^d\b{k}$. Using
\begin{align}
\int\frac{d^b\b{p}}{(2\pi)^d}\frac{1}{c^2k_0^2+\b{p}^2+r\mu^2}=-\frac{c^2k_0^2+r\mu^2}{8\pi^2\epsilon},
\end{align}
we find
\begin{align}
\delta Z_3^\text{(c)}=-\frac{\Delta}{8\pi^2\epsilon},\hspace{5mm}
\delta Z_r^\text{(c)}=-\frac{\Delta}{8\pi^2\epsilon}.
\end{align}

For Fig.~\ref{fig:diagrams}(b), we have
\begin{align}
\delta\Gamma_\text{div}^\text{(b)}&=-2Nh^2\sum_a\int\frac{d^Dk}{(2\pi)^D}\phi_a^*(k)\nn\\
&\hspace{5mm}\times\int\frac{d^Dp}{(2\pi)^D}\tr\frac{\slashed{p}(\slashed{p}+\slashed{k})}{p^2(p+k)^2}\phi_a(k),
\end{align}
where $\tr$ denotes a trace over spinor indices. Using Feynman parameters to express
\begin{align}\label{FeynParamx}
\frac{1}{p^2(p+k)^2}=\int_0^1\frac{dx}{[xp^2+(1-x)(p+k)^2]^2},
\end{align}
and shifting the integration variable $p\rightarrow p-(1-x)k$, we obtain
\begin{align}
\int\frac{d^Dp}{(2\pi)^D}\tr\frac{\slashed{p}(\slashed{p}+\slashed{k})}{p^2(p+k)^2}
=-\frac{k^2}{8\pi^2(\epsilon-\epsilon_\tau)},
\end{align}
using the fact that the gamma matrices are two-dimensional, as well as the 't Hooft-Veltman prescription~\cite{leibbrandt1975},
\begin{align}
\int\frac{d^Dp}{(2\pi)^D}\frac{1}{p^2}=0.
\end{align}
We thus obtain
\begin{align}
\delta Z_3^\text{(b)}=-\frac{Nh^2c^{-2}}{4\pi^2(\epsilon-\epsilon_\tau)},\hspace{5mm}
\delta Z_4^\text{(b)}=-\frac{Nh^2}{4\pi^2(\epsilon-\epsilon_\tau)}.
\end{align}

\subsection{Fermion two-point function}

A unique diagram, Fig.~\ref{fig:diagrams}(d), contributes to the renormalization of the fermion two-point function. The divergent part of the effective action is
\begin{align}
\delta\Gamma_\text{div}^\text{(d)}&=4h^2\sum_a\int\frac{d^Dk}{(2\pi)^D}\bar{\psi}_a(k)\nn\\
&\hspace{5mm}\times\int\frac{d^Dp}{(2\pi)^D}\frac{\slashed{p}+\slashed{k}}{(c^2p_0^2+\b{p}^2)(p+k)^2}\psi_a(k).
\end{align}
Using Feynman parameters as in Eq.~(\ref{FeynParamx}), and shifting $\b{p}\rightarrow\b{p}-(1-x)\b{k}$ to perform the integral over $\b{p}$ first, we have
\begin{align}
I_1&\equiv\int\frac{d^Dp}{(2\pi)^D}\frac{\slashed{p}+\slashed{k}}{(c^2p_0^2+\b{p}^2)(p+k)^2}\nn\\
&=\frac{\Gamma(\epsilon/2)}{(4\pi)^{d/2}}\int_0^1dx\int\frac{d^{\epsilon_\tau}p_0}{(2\pi)^{\epsilon_\tau}}\frac{\gamma_0(p_0+k_0)+x\boldsymbol{\gamma}\cdot\b{k}}{(M^2)^{\epsilon/2}},
\end{align}
where
\begin{align}
M^2&=\bigl(1+(c^2-1)x\bigr)\nn\\
&\hspace{5mm}\times\left[\ell_0^2+\frac{x(1-x)\b{k}^2}{1+(c^2-1)x}+\frac{x(1-x)c^2k_0^2}{\bigl(1+(c^2-1)x\bigr)^2}\right],
\end{align}
with
\begin{align}
\ell_0=p_0+\frac{(1-x)k_0}{1+(c^2-1)x}.
\end{align}
Shifting the integral over $p_0$ to one over $\ell_0$, we have, in the limit $\epsilon,\epsilon_\tau\rightarrow 0$,
\begin{align}
I_1&=\frac{1}{8\pi^2(\epsilon-\epsilon_\tau)}\int_0^1 dx\left(\frac{xc^2}{1+(c^2-1)x}\gamma_0k_0+x\boldsymbol{\gamma}\cdot\b{k}\right)\nn\\
&=\frac{1}{8\pi^2(\epsilon-\epsilon_\tau)}\left(\frac{c^2(c^2-1-\ln c^2)}{(c^2-1)^2}\gamma_0k_0+\frac{1}{2}\boldsymbol{\gamma}\cdot\b{k}\right).
\end{align}
We thus obtain
\begin{align}
\delta Z_1^\text{(d)}=-\frac{h^2 f(c^2)}{2\pi^2(\epsilon-\epsilon_\tau)},
\hspace{5mm}
\delta Z_2^\text{(d)}=-\frac{h^2}{4\pi^2(\epsilon-\epsilon_\tau)},
\end{align}
with $f(c^2)$ defined in Eq.~(\ref{f}).

\subsection{Boson self-interaction}

The relevant diagrams are given in Fig.~\ref{fig:diagrams}(e,f,g), where (e) and (g) are meant to include diagrams in all three ($s,t,u$) scattering channels.

For Fig.~\ref{fig:diagrams}(e), we have
\begin{align}
\delta\Gamma_\text{div}^\text{(e)}&=-2\lambda^4\sum_a\int\frac{d^Dk}{(2\pi)^D}\left(4|\phi_a|^2_{-k}|\phi_a|^2_k+(\phi_a^{*2})_{-k}(\phi_a^2)_k\right)\nn\\
&\hspace{5mm}\times\int\frac{d^Dp}{(2\pi)^D}\frac{1}{(c^2p_0^2+\b{p}^2)(c^2(p_0+k_0)^2+(\b{p}+\b{k})^2)}.
\end{align}
As before, we use Feynman parameters to perform the integral over $\b{p}$ first, shifting $\b{p}\rightarrow\b{p}-(1-x)\b{k}$,
\begin{align}\label{I2}
I_2&\equiv\int\frac{d^Dp}{(2\pi)^D}\frac{1}{(c^2p_0^2+\b{p}^2)(c^2(p_0+k_0)^2+(\b{p}+\b{k})^2)}\nn\\
&=\frac{\Gamma(\epsilon/2)}{(4\pi)^{d/2}}\int_0^1dx\int\frac{d^{\epsilon_\tau}\ell_0}{(2\pi)^{\epsilon_\tau}}
\frac{1}{(c^2\ell_0^2+Q^2)^{\epsilon/2}},
\end{align}
with $Q^2=x(1-x)(c^2k_0^2+\b{k}^2)$, and we have shifted the integral over $p_0$ to one over $\ell_0=p_0+(1-x)k_0$. Performing the integrals over $\ell_0$ and $x$, we obtain $I_2=1/[8\pi^2(\epsilon-\epsilon_\tau)]$, and thus
\begin{align}
\delta Z_5^\text{(e)}=\frac{5\lambda^2}{4\pi^2(\epsilon-\epsilon_\tau)}.
\end{align}

For Fig.~\ref{fig:diagrams}(f), we have
\begin{align}
\delta\Gamma_\text{div}^\text{(f)}&=4Nh^4\left(\prod_{i=1}^4\int\frac{d^Dk_i}{(2\pi)^D}\right)(2\pi)^D\delta\left(\sum_{i=1}^4k_i\right)\nn\\
&\hspace{5mm}\times\phi^*_a(-k_1)\phi(k_2)\phi^*_a(-k_3)\phi_a(k_4)\int\frac{d^Dp}{(2\pi)^D}\nn\\
&\hspace{5mm}\times\tr\frac{\slashed{p}(\slashed{p}-\slashed{k}_1)(\slashed{p}-\slashed{k}_1-\slashed{k}_2)(\slashed{p}+\slashed{k}_4)}{p^2(p-k_1)^2(p-k_1-k_2)^2(p+k_4)^2}.
\end{align}
Using four Feynman parameters,
\begin{align}
\frac{1}{A_1A_2A_3A_4}&=3!\int_0^1dx\int_0^1dy\int_0^1dz\int_0^1dw\nn\\
&\hspace{5mm}\times\frac{\delta(x+y+z+w-1)}{(xA_1+yA_2+zA_3+wA_4)^4},
\end{align}
as well as
\begin{align}
\tr\gamma_\mu\gamma_\lambda\gamma_\nu\gamma_\rho=2(\delta_{\mu\lambda}\delta_{\nu\rho}
+\delta_{\lambda\nu}\delta_{\mu\rho}-\delta_{\mu\nu}\delta_{\lambda\rho}),
\end{align}
to perform the spinor trace, we find that after shifting $p$ appropriately the denominator can be expressed as $(p^2+P^2)^4$ where $P^2$ is independent of $p$, and the numerator contains powers of $p$ ranging from one to four. For $D=4-(\epsilon-\epsilon_\tau)$, only the term with $(p^2)^2$ will give a pole in $\epsilon-\epsilon_\tau$. Using
\begin{align}
\int_0^1dx\int_0^1dy\int_0^1dz\int_0^1dw\,\delta(x+y+z+w-1)=\frac{1}{3!},
\end{align}
we find
\begin{align}
\delta\Gamma_\text{div}^\text{(f)}=\frac{Nh^4}{\pi^2(\epsilon-\epsilon_\tau)}\sum_a\int d^Dx\,|\phi_a|^4,
\end{align}
and thus
\begin{align}
\delta Z_5^\text{(f)}=-\frac{Nh^4\lambda^{-2}}{\pi^2(\epsilon-\epsilon_\tau)}.
\end{align}

The diagrams with one disorder vertex and one boson self-interaction vertex contribute to the renormalization of both $\lambda^2$ [Fig.~\ref{fig:diagrams}(g)] and $\Delta$ [Fig.~\ref{fig:diagrams}(h)]. Here we focus only on those diagrams that contribute to the renormalization of $\lambda^2$. We have
\begin{align}
\delta\Gamma_\text{div}^\text{(g)}&=2\lambda^2\Delta\sum_a\int\frac{d^Dk}{(2\pi)^D}\int\frac{d^Dq}{(2\pi)^D}
\nn\\
&\hspace{5mm}\times\Bigl(|\phi_a|^2_{-k}\phi_a^\alpha(k+q)\phi_a^\alpha(-q)\nn\\
&\hspace{5mm}+2(\phi_a^\alpha\phi_a^\beta)_{-k}\phi_a^\alpha(k+q)\phi_a^\beta(-q)\Bigr)\nn\\
&\hspace{5mm}\times\int\frac{d^d\b{p}}{(2\pi)^d}\frac{1}{(c^2q_0^2+\b{p}^2)(c^2(q_0+k_0)^2+(\b{p}+\b{k})^2)},
\end{align}
denoting $\phi^1_a=\re\phi_a$, $\phi^2_a=\im\phi_a$, and with sums over repeated indices $\alpha,\beta=1,2$ understood. Denoting $m_1^2=c^2q_0^2$ and $m_2^2=c^2(q_0+k_0)^2$, the loop integral is
\begin{align}\label{LoopIntegral}
\int\frac{d^d\b{p}}{(2\pi)^d}\frac{1}{(\b{p}^2+m_1^2)((\b{p}+\b{k})^2+m_2^2)}=\frac{1}{8\pi^2\epsilon},
\end{align}
using Feynman parameters and shifting $\b{p}\rightarrow\b{p}-(1-x)\b{k}$. We thus obtain
\begin{align}
\delta\Gamma_\text{div}^\text{(g)}=\frac{3\lambda^2\Delta}{4\pi^2\epsilon}\sum_a\int d^Dx\,|\phi_a|^4,
\end{align}
and
\begin{align}
\delta Z_5^\text{(g)}=-\frac{3\Delta}{4\pi^2\epsilon}.
\end{align}

\subsection{Disorder strength}

The two diagrams are Fig.~\ref{fig:diagrams}(h) and (i). For Fig.~\ref{fig:diagrams}(h), we have
\begin{align}
\delta\Gamma_\text{div}^\text{(h)}&=4\lambda^2\Delta\sum_{ab}\int\frac{d^d\b{k}}{(2\pi)^d}\int d^{\epsilon_\tau}\tau\int d^{\epsilon_\tau}\tau'|\phi_a|^2_{-\b{k},\tau}|\phi_b|^2_{\b{k},\tau'}\nn\\
&\hspace{5mm}\times\int\frac{d^Dp}{(2\pi)^D}\frac{1}{(c^2p_0^2+\b{p}^2)(c^2p_0^2+(\b{p}+\b{k})^2)}.
\end{align}
The loop integral is the same as $I_2$ in Eq.~(\ref{I2}), but with $k_0=0$, which does not change the result $I_2=1/[8\pi^2(\epsilon-\epsilon_\tau)]$ in the limit $\epsilon,\epsilon_\tau\rightarrow 0$. We thus have
\begin{align}
\delta\Gamma_\text{div}^\text{(h)}=\frac{\lambda^2\Delta}{2\pi^2(\epsilon-\epsilon_\tau)}\sum_{ab}\int d^d\b{x}\,d^{\epsilon_\tau}\tau\,d^{\epsilon_\tau}\tau'\,|\phi_a|^2_{\b{x},\tau}|\phi_b|^2_{\b{x},\tau'},
\end{align}
hence
\begin{align}
\delta Z_7^\text{(h)}=\frac{\lambda^2}{\pi^2(\epsilon-\epsilon_\tau)}.
\end{align}

Finally, ignoring a term which vanishes in the replica limit, Fig.~\ref{fig:diagrams}(i) is given by the sum of two contributions:
\begin{align}
\delta\Gamma_\text{div}^\text{(i,1)}&=-\Delta^2\sum_{ab}\int\frac{d^d\b{k}}{(2\pi)^d}\int d^{\epsilon_\tau}\tau\int\frac{d^Dq}{(2\pi)^D}\nn\\
&\hspace{5mm}\times|\phi_a|^2_{-\b{k},\tau}\phi_b^\alpha(\b{k}+\b{q},q_0)\phi_b^\alpha(-q)\nn\\
&\hspace{5mm}\times\int\frac{d^d\b{p}}{(2\pi)^d}\frac{1}{(c^2q_0^2+\b{p}^2)(c^2q_0^2+(\b{p}+\b{k})^2)},
\end{align}
and
\begin{align}
\delta\Gamma_\text{div}^\text{(i,2)}&=-\Delta^2\sum_{ab}\left(\prod_{i=1}^4\int\frac{d^d\b{k}_i}{(2\pi)^d}\right)(2\pi)^d\delta\left(\sum_{i=1}^4\b{k}_i\right)\nn\\
&\hspace{5mm}\times\int\frac{d^{\epsilon_\tau}p_0}{(2\pi)^{\epsilon_\tau}}\int\frac{d^{\epsilon_\tau}q_0}{(2\pi)^{\epsilon_\tau}}\nn\\
&\hspace{5mm}\times\phi_a^\alpha(\b{k}_1,p_0)\phi_a^\alpha(\b{k}_4,-p_0)\phi_b^\beta(\b{k}_3,q_0)\phi_b^\beta(\b{k}_2,-q_0)\nn\\
&\hspace{5mm}\times\int\frac{d^d\b{p}}{(2\pi)^d}\frac{1}{(c^2p_0^2+\b{p}^2)(c^2q_0^2+(\b{p}+\b{k}_3+\b{k}_4)^2)}.
\end{align}
Both integrals over the loop momentum $\b{p}$ are of the form (\ref{LoopIntegral}), and thus evaluate to $1/(8\pi^2\epsilon)$. Performing the remaining integrals, we obtain
\begin{align}
\delta\Gamma_\text{div}^\text{(i,1)}+\delta\Gamma_\text{div}^\text{(i,2)}&=-\frac{\Delta^2}{4\pi^2\epsilon}\sum_{ab}\int d^d\b{x}\,d^{\epsilon_\tau}\tau\,d^{\epsilon_\tau}\tau'\nn\\
&\hspace{5mm}\times|\phi_a|^2(\b{x},\tau)|\phi_b|^2(\b{x},\tau'),
\end{align}
thus
\begin{align}
\delta Z_7^\text{(i)}=-\frac{\Delta}{2\pi^2\epsilon}.
\end{align}

Adding up the various contributions and rescaling the couplings $\lambda^2$, $h^2$, and $\Delta$ by $(4\pi)^2$, we obtain the renormalization constants in Eq.~(\ref{Z1})-(\ref{Zr}).

\section{Oscillatory corrections to scaling}
\label{app:oscillatory}

We derive the existence of oscillatory corrections to scaling~\cite{khmelnitskii1978} for $N\geq 7$ at the DFP 2 due to the presence of a pair of complex-conjugate eigenvalues of the stability matrix. Passing over to a Wilsonian description, and ignoring corrections to the dynamic critical exponent, the two-point function of the order parameter $\chi(\b{q})=\langle\phi(\b{q})\phi^*(\b{q})\rangle$ obeys the scaling relation $\chi(\b{q},r(0))=e^{(2-\eta_\phi)\ell}\chi(e^\ell\b{q},r(\ell))$, where $\ell$ is an infrared scale parameter, $r(0)$ is the bare relevant tuning parameter for the transition, and $r(\ell)$ is the renormalized tuning parameter, which obeys the differential equation
\begin{align}\label{drdl}
\frac{dr(\ell)}{d\ell}=[2-\gamma_{m^2}(\b{g}(\ell))]r(\ell).
\end{align}
Similarly, $\b{g}(\ell)=\bigl(c^2,h^2,\lambda^2,\Delta)$ is a vector of renormalized couplings, which obeys the differential equation
\begin{align}\label{dgdl}
\frac{d\b{g}(\ell)}{d\ell}=\boldsymbol{\beta}(\b{g}(\ell)),
\end{align}
where $\boldsymbol{\beta}=(\beta_{c^2},\beta_{h^2},\beta_{\lambda^2},\beta_\Delta)$ is a vector of beta functions given by Eq.~(\ref{bc2})-(\ref{bD}), but with a minus sign since $d\ell=-d\ln\mu$. Defining $\ell_r$ such that $r(\ell_r)=r_0$ for some arbitrary constant $r_0$, we find that the uniform thermodynamic susceptibility behaves as $\chi(\b{q}=0,r)\sim e^{(2-\eta_\phi)\ell_r}$ where we now denote $r(0)$ by $r$ for simplicity, and $\ell_r$ depends on $r$ in a manner to be determined. Integrating Eq.~(\ref{drdl}) from $\ell=0$ to $\ell=\ell_r$, we find
\begin{align}\label{logr0r}
\ln\left(\frac{r_0}{r}\right)=\int_0^{\ell_r}d\ell\,[2-\gamma_{m^2}(\b{g}(\ell))].
\end{align}
Linearizing Eq.~(\ref{dgdl}) near the fixed point $\b{g}_*$, we have
\begin{align}\label{linearized}
\frac{d}{d\ell}\bigl(\b{g}(\ell)-\b{g}_*\bigr)=M\bigl(\b{g}(\ell)-\b{g}_*\bigr),
\end{align}
which is solved by diagonalizing $M=PDP^{-1}$ where $D$ is a diagonal matrix. Now, $\gamma_{m^2}$ in Eq.~(\ref{logr0r}) can be read off from Eq.~(\ref{nuinv}), and is linear in the couplings:
\begin{align}
\gamma_{m^2}(\b{g}(\ell))=\b{a}\cdot\b{g}(\ell)
=\b{a}\cdot\b{g}_*+\sum_iu_i(0)\b{a}\cdot\b{v}_i e^{-\omega_i\ell},
\end{align}
where the eigenvalues of $M$ are denoted as $-\omega_i$, $\b{v}_i$ are the respective eigenvectors, and $\b{u}(0)$ is a vector of initial conditions,
\begin{align}\label{u0}
\b{u}(0)=P^{-1}\bigl(\b{g}(0)-\b{g}_*\bigr).
\end{align}
Substituting into Eq.~(\ref{logr0r}), we obtain
\begin{align}\label{logr0rbis}
\ln\left(\frac{r_0}{r}\right)=\nu^{-1}\ell_r+\sum_i\frac{u_i(0)}{\omega_i}\b{a}\cdot\b{v}_i\left(e^{-\omega_i\ell_r}-1\right),
\end{align}
where $\nu^{-1}=2-\gamma_{m^2}(\b{g}_*)$. Assuming that the deviation (\ref{u0}) from the fixed point is small, we can solve for $\ell_r$ to $\mathcal{O}\bigl(\b{u}(0)\bigr)$,
\begin{align}
\ell_r&=\nu\ln\left(\frac{r_0}{r}\right)-\sum_i\frac{\nu u_i(0)}{\omega_i}\b{a}\cdot\b{v}_i\left[\left(\frac{r}{r_0}\right)^{\nu\omega_i}-1\right]\nn\\
&\phantom{=}+\mathcal{O}\bigl(\b{u}(0)^2\bigr).
\end{align}
The susceptibility thus becomes
\begin{align}
\chi\sim |r|^{-\gamma}\left[1-\sum_i\frac{\gamma u_i(0)}{\omega_i}\b{a}\cdot\b{v}_i\left(\frac{r}{r_0}\right)^{\nu\omega_i}+\mathcal{O}\bigl(\b{u}(0)^2\bigr)\right],
\end{align}
where $\gamma=(2-\eta_\phi)\nu$ is the usual susceptibility exponent.

Real (positive) eigenvalues $\omega\in\mathbb{R}$ produce the usual corrections to scaling $\chi\sim|r|^{-\gamma}(1+C|r|^{\nu\omega}+\ldots)$~\cite{Goldenfeld}. Since the stability matrix $M$ in Eq.~(\ref{linearized}) is real, complex eigenvalues $\omega=\omega'+i\omega''$, if any, must come in complex-conjugate pairs $\omega,\omega^*$. The associated eigenvectors $\b{v},\b{v}^*$ are also complex conjugates since $M\b{v}=-\omega\b{v}$ and $M$ is real. Finally, since the components $u_i$ obey the differential equation $du_i/d\ell=-\omega_iu_i$, the component of $\b{u}(0)$ associated with $\omega^*$ must also be the complex conjugate of the component associated with $\omega$. As a result the corrections to scaling due to a single pair of complex-conjugate eigenvalues $\omega'\pm i\omega''$ are of the form
\begin{align}
\chi&\sim|r|^{-\gamma}\left[1+\left(\frac{1}{2}Ce^{i\phi}\left(\frac{r}{r_0}\right)^{\nu(\omega'+i\omega'')}+\text{c.c.}\right)+\ldots\right]\nn\\
&\sim|r|^{-\gamma}\left[1+C\left|\frac{r}{r_0}\right|^{\nu\omega'}\cos\left(\nu\omega''\ln\left|\frac{r}{r_0}\right|+\phi\right)+\ldots\right],
\end{align}
where $C$ and $\phi$ are nonuniversal constants, but the exponents $\omega'$ and $\omega''$ (see Fig.~\ref{fig:corrscaling}) are universal.

\bibliography{DisorderQCP}

\begin{thebibliography}{104}
\bibitem [{Not()}]{NotezBC}%
  \BibitemOpen
  \href@noop {} {}\bibinfo {note} {Note that naively substituting $h_*^2=0$
  into Eq.~(\ref{z}) would imply that this fixed point has $z=1$, which is
  incorrect. The issue is that this fixed point has $c_*^2=0$, but the
  renormalization constants (\ref{Z1})-(\ref{Zr}) are calculated assuming
  $c^2\neq 0$. To calculate $z$ at the bosonic disordered fixed point one
  should rescale fields and redefine couplings in Eq.~(\ref{Sreplica}) in such
  a way as to eliminate the parameter $c^2$ in front of
  $|\partial_\tau\phi_a|^2$ at the expense of multiplying
  $i\bar{\psi}_a\slashed{\partial}_\tau\psi_a$ by $1/c$. The dynamic critical
  exponent is then given by
  $z=1+\frac{1}{2}(\gamma_3-\gamma_4)=1+\frac{3\epsilon_\tau}{4}$, in agreement
  with known results (see, e.g., Ref.~\cite{QPT}).}\BibitemShut {Stop}%
\end{thebibliography}

\end{document}